\documentclass[reqno,12pt,a4paper]{amsart}

\usepackage{mathrsfs}
\usepackage{amssymb}
\usepackage{amsfonts}
\usepackage{latexsym}
\usepackage{amsthm}

\usepackage{graphicx}
\def\lb{\label}

\newcommand{\er}[1]{\textrm{(\ref{#1})}}

\begin{document}


\renewcommand{\theequation}{\arabic{section}.\arabic{equation}}
\theoremstyle{plain}
\newtheorem{theorem}{\bf Theorem}[section]
\newtheorem{lemma}[theorem]{\bf Lemma}
\newtheorem{corollary}[theorem]{\bf Corollary}
\newtheorem{proposition}[theorem]{\bf Proposition}
\newtheorem{definition}[theorem]{\bf Definition}
\newtheorem{remark}[theorem]{\it Remark}
\newtheorem{condition}[theorem]{\bf Condition}

\def\a{\alpha}  \def\cA{{\mathcal A}}     \def\bA{{\bf A}}  \def\mA{{\mathscr A}}
\def\b{\beta}   \def\cB{{\mathcal B}}     \def\bB{{\bf B}}  \def\mB{{\mathscr B}}
\def\g{\gamma}  \def\cC{{\mathcal C}}     \def\bC{{\bf C}}  \def\mC{{\mathscr C}}
\def\G{\Gamma}  \def\cD{{\mathcal D}}     \def\bD{{\bf D}}  \def\mD{{\mathscr D}}
\def\d{\delta}  \def\cE{{\mathcal E}}     \def\bE{{\bf E}}  \def\mE{{\mathscr E}}
\def\D{\Delta}  \def\cF{{\mathcal F}}     \def\bF{{\bf F}}  \def\mF{{\mathscr F}}
\def\c{\chi}    \def\cG{{\mathcal G}}     \def\bG{{\bf G}}  \def\mG{{\mathscr G}}
\def\z{\zeta}   \def\cH{{\mathcal H}}     \def\bH{{\bf H}}  \def\mH{{\mathscr H}}
\def\e{\eta}    \def\cI{{\mathcal I}}     \def\bI{{\bf I}}  \def\mI{{\mathscr I}}
\def\p{\psi}    \def\cJ{{\mathcal J}}     \def\bJ{{\bf J}}  \def\mJ{{\mathscr J}}
\def\vT{\Theta} \def\cK{{\mathcal K}}     \def\bK{{\bf K}}  \def\mK{{\mathscr K}}
\def\k{\kappa}  \def\cL{{\mathcal L}}     \def\bL{{\bf L}}  \def\mL{{\mathscr L}}
\def\l{\lambda} \def\cM{{\mathcal M}}     \def\bM{{\bf M}}  \def\mM{{\mathscr M}}
\def\L{\Lambda} \def\cN{{\mathcal N}}     \def\bN{{\bf N}}  \def\mN{{\mathscr N}}
\def\m{\mu}     \def\cO{{\mathcal O}}     \def\bO{{\bf O}}  \def\mO{{\mathscr O}}
\def\n{\nu}     \def\cP{{\mathcal P}}     \def\bP{{\bf P}}  \def\mP{{\mathscr P}}
\def\r{\rho}    \def\cQ{{\mathcal Q}}     \def\bQ{{\bf Q}}  \def\mQ{{\mathscr Q}}
\def\s{\sigma}  \def\cR{{\mathcal R}}     \def\bR{{\bf R}}  \def\mR{{\mathscr R}}
\def\S{\Sigma}  \def\cS{{\mathcal S}}     \def\bS{{\bf S}}  \def\mS{{\mathscr S}}
\def\t{\tau}    \def\cT{{\mathcal T}}     \def\bT{{\bf T}}  \def\mT{{\mathscr T}}
\def\f{\phi}    \def\cU{{\mathcal U}}     \def\bU{{\bf U}}  \def\mU{{\mathscr U}}
\def\F{\Phi}    \def\cV{{\mathcal V}}     \def\bV{{\bf V}}  \def\mV{{\mathscr V}}
\def\P{\Psi}    \def\cW{{\mathcal W}}     \def\bW{{\bf W}}  \def\mW{{\mathscr W}}
\def\o{\omega}  \def\cX{{\mathcal X}}     \def\bX{{\bf X}}  \def\mX{{\mathscr X}}
\def\x{\xi}     \def\cY{{\mathcal Y}}     \def\bY{{\bf Y}}  \def\mY{{\mathscr Y}}
\def\X{\Xi}     \def\cZ{{\mathcal Z}}     \def\bZ{{\bf Z}}  \def\mZ{{\mathscr Z}}
\def\O{\Omega}
\def\th{\theta}
\def\vs{\varsigma}

\newcommand{\gA}{\mathfrak{A}}
\newcommand{\gB}{\mathfrak{B}}
\newcommand{\gC}{\mathfrak{C}}
\newcommand{\gD}{\mathfrak{D}}
\newcommand{\gE}{\mathfrak{E}}
\newcommand{\gF}{\mathfrak{F}}
\newcommand{\gG}{\mathfrak{G}}
\newcommand{\gH}{\mathfrak{H}}
\newcommand{\gI}{\mathfrak{I}}
\newcommand{\gJ}{\mathfrak{J}}
\newcommand{\gK}{\mathfrak{K}}
\newcommand{\gL}{\mathfrak{L}}
\newcommand{\gM}{\mathfrak{M}}
\newcommand{\gN}{\mathfrak{N}}
\newcommand{\gO}{\mathfrak{O}}
\newcommand{\gP}{\mathfrak{P}}
\newcommand{\gQ}{\mathfrak{Q}}
\newcommand{\gR}{\mathfrak{R}}
\newcommand{\gS}{\mathfrak{S}}
\newcommand{\gT}{\mathfrak{T}}
\newcommand{\gU}{\mathfrak{U}}
\newcommand{\gV}{\mathfrak{V}}
\newcommand{\gW}{\mathfrak{W}}
\newcommand{\gX}{\mathfrak{X}}
\newcommand{\gY}{\mathfrak{Y}}
\newcommand{\gZ}{\mathfrak{Z}}

\def\ve{\varepsilon}   \def\vt{\vartheta}    \def\vp{\varphi}    \def\vk{\varkappa}

\def\Z{{\mathbb Z}}    \def\R{{\mathbb R}}   \def\C{{\mathbb C}}    \def\K{{\mathbb K}}
\def\T{{\mathbb T}}    \def\N{{\mathbb N}}   \def\dD{{\mathbb D}}


\def\la{\leftarrow}              \def\ra{\rightarrow}            \def\Ra{\Rightarrow}
\def\ua{\uparrow}                \def\da{\downarrow}
\def\lra{\leftrightarrow}        \def\Lra{\Leftrightarrow}


\def\lt{\biggl}                  \def\rt{\biggr}
\def\ol{\overline}               \def\wt{\widetilde}
\def\no{\noindent}


\let\ge\geqslant                 \let\le\leqslant
\def\lan{\langle}                \def\ran{\rangle}
\def\/{\over}                    \def\iy{\infty}
\def\sm{\setminus}               \def\es{\emptyset}
\def\ss{\subset}                 \def\ts{\times}
\def\pa{\partial}                \def\os{\oplus}
\def\om{\ominus}                 \def\ev{\equiv}
\def\iint{\int\!\!\!\int}        \def\iintt{\mathop{\int\!\!\int\!\!\dots\!\!\int}\limits}
\def\el2{\ell^{\,2}}             \def\1{1\!\!1}
\def\sh{\sharp}
\def\wh{\widehat}
\def\bs{\backslash}

\def\sh{\mathop{\mathrm{sh}}\nolimits}
\def\Area{\mathop{\mathrm{Area}}\nolimits}
\def\arg{\mathop{\mathrm{arg}}\nolimits}
\def\const{\mathop{\mathrm{const}}\nolimits}
\def\det{\mathop{\mathrm{det}}\nolimits}
\def\diag{\mathop{\mathrm{diag}}\nolimits}
\def\diam{\mathop{\mathrm{diam}}\nolimits}
\def\dim{\mathop{\mathrm{dim}}\nolimits}
\def\dist{\mathop{\mathrm{dist}}\nolimits}
\def\Im{\mathop{\mathrm{Im}}\nolimits}
\def\Iso{\mathop{\mathrm{Iso}}\nolimits}
\def\Ker{\mathop{\mathrm{Ker}}\nolimits}
\def\Lip{\mathop{\mathrm{Lip}}\nolimits}
\def\rank{\mathop{\mathrm{rank}}\limits}
\def\Ran{\mathop{\mathrm{Ran}}\nolimits}
\def\Re{\mathop{\mathrm{Re}}\nolimits}
\def\Res{\mathop{\mathrm{Res}}\nolimits}
\def\res{\mathop{\mathrm{res}}\limits}
\def\sign{\mathop{\mathrm{sign}}\nolimits}
\def\span{\mathop{\mathrm{span}}\nolimits}
\def\supp{\mathop{\mathrm{supp}}\nolimits}
\def\Tr{\mathop{\mathrm{Tr}}\nolimits}
\def\BBox{\hspace{1mm}\vrule height6pt width5.5pt depth0pt \hspace{6pt}}
\def\as{\text{as}}
\def\all{\text{all}}
\def\where{\text{where}}
\def\Dom{\mathop{\mathrm{Dom}}\nolimits}


\newcommand\nh[2]{\widehat{#1}\vphantom{#1}^{(#2)}}
\def\dia{\diamond}

\def\Oplus{\bigoplus\nolimits}



\def\qqq{\qquad}
\def\qq{\quad}
\let\ge\geqslant
\let\le\leqslant
\let\geq\geqslant
\let\leq\leqslant
\newcommand{\ca}{\begin{cases}}
\newcommand{\ac}{\end{cases}}
\newcommand{\ma}{\begin{pmatrix}}
\newcommand{\am}{\end{pmatrix}}
\renewcommand{\[}{\begin{equation}}
\renewcommand{\]}{\end{equation}}
\def\eq{\begin{equation}}
\def\qe{\end{equation}}
\def\[{\begin{equation}}
\def\bu{\bullet}

\newcommand{\fr}{\frac}
\newcommand{\tf}{\tfrac}

\title[Third order operators with three-point conditions]
{Third order operators with three-point conditions
associated with Boussinesq's equation}

\date{\today}
\author[Andrey Badanin]{Andrey Badanin}
\author[Evgeny Korotyaev]{Evgeny L. Korotyaev}
\address{Saint-Petersburg
State University, Universitetskaya nab. 7/9, St. Petersburg,
199034 Russia,
an.badanin@gmail.com,\  a.badanin@spbu.ru,\
korotyaev@gmail.com,\  e.korotyaev@spbu.ru}


\begin{abstract}
We consider a non-self-adjoint third order operator on the interval
$[0,2]$ with real 1-periodic coefficients and three-point Dirichlet
conditions at the points 0, 1 and 2. The eigenvalues of this
operator consist an auxiliary spectrum for the inverse spectral
problem associated with the good Boussinesq equation.
We determine eigenvalue asymptotics at high energy and
the trace formula for the operator.

\end{abstract}

\maketitle


\section {Introduction and main results}
\setcounter{equation}{0}

We consider a non-self-adjoint operator $H$ acting on $L^2(0,2)$ and
given by
\[
\lb{Hpq}
Hy=(y''+py)'+py'+qy,\qqq y(0)=y(1)=y(2)=0,
\]
where $p,q$ are real  1-periodic coefficients
$p,q\in L^1(\T),\T=\R/\Z$. The operator is defined on the domain
\[
\lb{cDH}
\begin{aligned}
\Dom(H)=\Big\{y\in L^2(0,2):(y''+py)'+py'+qy\in L^2(0,2),
\\
y'',(y''+py)'\in L^1(0,2),
 y(0)=y(1)=y(2)=0\Big\}.
\end{aligned}
\]

The multi-point problems for linear ordinary differential operators
are well known, see, e.g.,
the papers \cite{EHH92}, \cite{L68}, \cite{Po08} and references therein.
Papanicolaou \cite{Pa03}, \cite{Pa05} considered the non-linear equation
associated with the linear fourth order
Euler-Bernoulli operators on the circle and studied the multi-point problem
for this operator.
Trace formulas for multipoint problems for two-term $2n$-order differential operators
were obtained by Belabbasi \cite{Be83}.

The operator $H$ is used in the integration of the so-called
{\it good Boussinesq equation} on the circle,
$$
p_{tt}=-{1\/3}(p_{xxxx}+4(p^2)_{xx}), \qqq p_t=q_x,
$$
see \cite{McK81} and references therein. It is equivalent to the Lax
equation $L_t=LA-AL$, where the operators $L,A$ act on $L^2(\T)$ and
have the form $L=\pa^3+p\pa+\pa p+q,A=-\pa^2-{4\/3}p$.
Kalantarov and Ladyzhenskaja \cite{KL77} showed that
the good Boussinesq equation has blow-up solutions.
The spectrum
of the operator $H$ is an auxiliary spectrum for the Boussinesq
equation similar to the Dirichlet spectrum on the unit interval for
the Korteweg-de Vries equation on the circle. McKean \cite{McK81}
considered the operator $H$ with the coefficients $p,q\in
C^\iy(\T)$.
The auxiliary spectrum in the finite-gap case was discussed
by Dickson, Gesztesy and Unterkofler \cite{DGU99}, \cite{DGU99x}.
In addition, there are given some interesting exactly calculating examples.
Self-adjoint third order operators $i\pa^3+ip\pa+i\pa
p+q$ associated with the {\it bad Boussinesq equations} on the
circle was studied by Badanin and Korotyaev \cite{BK12},
\cite{BK14}. The inverse scattering theory for the self-adjoint
third order operator with decreasing coefficients was developed in
\cite{DTT82}. Korotyaev \cite{K16} considered resonances for
third-order operator.

Consider the differential equation
\[
\lb{1b}
y'''+(py)'+py'+q y=\l y,\qqq \l\in\C.
\]
Rewrite this equation in the vector form
$$
{\bf y}'=\cP{\bf y}, \qqq\where\qq {\bf y}=\ma y\\y'\\y''+py\am,
\qq \cP=\ma 0&1&0\\-p &0&1\\\l-q&-p&0\am.
$$
The matrix-valued solution $M(x,\l),(x,\l)\in\C\ts\R$, of the initial problem
\[
\lb{me1}
M'= \cP M,
\qqq M(0,\l)=\1_3,
\]
is called the fundamental matrix,
here and below $\1_3$ is the $3\ts 3$ identity matrix.
It has the form
\[
\lb{deM}
M=\ma\vp_1&\vp_2&\vp_3\\
\vp_1'&\vp_2'&\vp_3'\\
\vp_1''+p\vp_1&\vp_2''+p\vp_2&\vp_3''+p\vp_3\am,
\]
where $\vp_1, \vp_2, \vp_3$ are the fundamental solutions of equation
\er{1b} satisfying the initial conditions
\er{me1}.
Each matrix-valued function $M(x,\cdot),x\in\R$, is entire, real for $\l\in\R$
and satisfies the Liouville identity
$\det M(x,\l)=1$ for all $\l\in\C$.

It is well known that the spectrum $\s(H)$ of $H$ is pure discrete
and satisfies
\[
\lb{spec}
\s(H)=\{\l\in\C:D(\l)=0\}.
\]
Here $D$ is the entire function given by
\[
\lb{defsi}
D(\l)=\det\ma\vp_2(1,\l)&\vp_3(1,\l)\\
\vp_2(2,\l)&\vp_3(2,\l)\am,
\]
see, e.g., \cite{W18}, where Green's function is studied.
The spectrum consists of eigenvalues
$\m_n,n\in\Z\sm\{0\}$, labeled by
\[
\lb{labev}
...\le \Re\m_{-2}\le \Re\m_{-1}\le \Re\m_{1}\le \Re\m_{2}\le...
\]
counting with algebraic multiplicities. Note that some eigenvalues
may be non-real, see~Fig.~\ref{Figspec}, and we have no information
on how large the algebraic multiplicity of the eigenvalue can be.
\begin{figure}
\centering
\tiny
\unitlength 0.9mm
\linethickness{0.2pt}
\ifx\plotpoint\undefined\newsavebox{\plotpoint}\fi 
\begin{picture}(105.25,52)(0,0)
\put(2.5,26){\line(1,0){102.75}}
\put(53.25,52){\line(0,-1){49}}
\put(63.25,32){\circle*{1.414}}
\put(63.25,19.5){\circle*{1.414}}
\put(79.25,26){\circle*{1.414}}
\put(94,26){\circle*{1.414}}
\put(25.5,26){\circle*{1.414}}
\put(7.75,26){\circle*{1.414}}
\put(42.25,36){\circle*{1.414}}
\put(42.25,15.5){\circle*{1.414}}
\put(98,38.5){\makebox(0,0)[cc]{$\l$}}
\put(98,38.5){\circle{6.403}}
\put(65,34.5){\makebox(0,0)[cc]{$\m_1$}}
\put(65,16.5){\makebox(0,0)[cc]{$\m_2$}}
\put(79,23.25){\makebox(0,0)[cc]{$\m_3$}}
\put(94.25,23.75){\makebox(0,0)[cc]{$\m_4$}}
\put(42,38.75){\makebox(0,0)[cc]{$\m_{-1}$}}
\put(41.75,12.5){\makebox(0,0)[cc]{$\m_{-2}$}}
\put(24.75,29.25){\makebox(0,0)[cc]{$\m_{-3}$}}
\put(6.75,29.25){\makebox(0,0)[cc]{$\m_{-4}$}}
\put(51.25,24){\makebox(0,0)[cc]{$0$}}
\put(74.158,26){\line(0,1){.9529}}
\put(74.136,26.953){\line(0,1){.9509}}
\put(74.071,27.904){\line(0,1){.9469}}
\multiput(73.962,28.851)(-.030289,.188203){5}{\line(0,1){.188203}}
\multiput(73.811,29.792)(-.032362,.155522){6}{\line(0,1){.155522}}
\multiput(73.617,30.725)(-.029563,.115414){8}{\line(0,1){.115414}}
\multiput(73.38,31.648)(-.030926,.101286){9}{\line(0,1){.101286}}
\multiput(73.102,32.56)(-.031959,.089794){10}{\line(0,1){.089794}}
\multiput(72.782,33.458)(-.032744,.080222){11}{\line(0,1){.080222}}
\multiput(72.422,34.34)(-.033336,.072093){12}{\line(0,1){.072093}}
\multiput(72.022,35.205)(-.0313601,.0604272){14}{\line(0,1){.0604272}}
\multiput(71.583,36.051)(-.0318094,.0550062){15}{\line(0,1){.0550062}}
\multiput(71.106,36.876)(-.0321406,.0501556){16}{\line(0,1){.0501556}}
\multiput(70.592,37.679)(-.0323699,.0457775){17}{\line(0,1){.0457775}}
\multiput(70.041,38.457)(-.0325103,.0417961){18}{\line(0,1){.0417961}}
\multiput(69.456,39.209)(-.0325718,.0381515){19}{\line(0,1){.0381515}}
\multiput(68.837,39.934)(-.0325629,.034796){20}{\line(0,1){.034796}}
\multiput(68.186,40.63)(-.0341149,.0332757){20}{\line(-1,0){.0341149}}
\multiput(67.504,41.296)(-.0374695,.0333541){19}{\line(-1,0){.0374695}}
\multiput(66.792,41.929)(-.0411146,.0333679){18}{\line(-1,0){.0411146}}
\multiput(66.052,42.53)(-.0450981,.03331){17}{\line(-1,0){.0450981}}
\multiput(65.285,43.096)(-.04948,.0331712){16}{\line(-1,0){.04948}}
\multiput(64.493,43.627)(-.0543364,.0329405){15}{\line(-1,0){.0543364}}
\multiput(63.678,44.121)(-.0597656,.0326034){14}{\line(-1,0){.0597656}}
\multiput(62.842,44.578)(-.0658963,.0321415){13}{\line(-1,0){.0658963}}
\multiput(61.985,44.995)(-.0729,.03153){12}{\line(-1,0){.0729}}
\multiput(61.11,45.374)(-.081013,.030736){11}{\line(-1,0){.081013}}
\multiput(60.219,45.712)(-.100625,.033015){9}{\line(-1,0){.100625}}
\multiput(59.313,46.009)(-.114778,.031944){8}{\line(-1,0){.114778}}
\multiput(58.395,46.265)(-.132702,.030491){7}{\line(-1,0){.132702}}
\multiput(57.466,46.478)(-.15628,.02848){6}{\line(-1,0){.15628}}
\multiput(56.529,46.649)(-.23612,.03199){4}{\line(-1,0){.23612}}
\put(55.584,46.777){\line(-1,0){.9493}}
\put(54.635,46.862){\line(-1,0){.9522}}
\put(53.683,46.903){\line(-1,0){.9531}}
\put(52.729,46.901){\line(-1,0){.952}}
\put(51.777,46.856){\line(-1,0){.949}}
\multiput(50.828,46.767)(-.23599,-.03299){4}{\line(-1,0){.23599}}
\multiput(49.884,46.635)(-.156158,-.029138){6}{\line(-1,0){.156158}}
\multiput(48.947,46.46)(-.132573,-.03105){7}{\line(-1,0){.132573}}
\multiput(48.019,46.243)(-.114642,-.032427){8}{\line(-1,0){.114642}}
\multiput(47.102,45.983)(-.100485,-.033439){9}{\line(-1,0){.100485}}
\multiput(46.198,45.682)(-.080882,-.031078){11}{\line(-1,0){.080882}}
\multiput(45.308,45.34)(-.072767,-.031837){12}{\line(-1,0){.072767}}
\multiput(44.435,44.958)(-.0657602,-.032419){13}{\line(-1,0){.0657602}}
\multiput(43.58,44.537)(-.0596276,-.032855){14}{\line(-1,0){.0596276}}
\multiput(42.745,44.077)(-.054197,-.0331692){15}{\line(-1,0){.054197}}
\multiput(41.932,43.579)(-.0493397,-.0333795){16}{\line(-1,0){.0493397}}
\multiput(41.143,43.045)(-.0449573,-.0334998){17}{\line(-1,0){.0449573}}
\multiput(40.379,42.476)(-.0409736,-.0335409){18}{\line(-1,0){.0409736}}
\multiput(39.641,41.872)(-.0373286,-.0335117){19}{\line(-1,0){.0373286}}
\multiput(38.932,41.235)(-.0339743,-.0334193){20}{\line(-1,0){.0339743}}
\multiput(38.252,40.567)(-.0324159,-.0349329){20}{\line(0,-1){.0349329}}
\multiput(37.604,39.868)(-.0324107,-.0382884){19}{\line(0,-1){.0382884}}
\multiput(36.988,39.141)(-.0323338,-.0419328){18}{\line(0,-1){.0419328}}
\multiput(36.406,38.386)(-.0321766,-.0459136){17}{\line(0,-1){.0459136}}
\multiput(35.859,37.606)(-.0319288,-.0502906){16}{\line(0,-1){.0502906}}
\multiput(35.348,36.801)(-.0315772,-.0551398){15}{\line(0,-1){.0551398}}
\multiput(34.875,35.974)(-.0334977,-.0652173){13}{\line(0,-1){.0652173}}
\multiput(34.439,35.126)(-.033031,-.072233){12}{\line(0,-1){.072233}}
\multiput(34.043,34.259)(-.032406,-.08036){11}{\line(0,-1){.08036}}
\multiput(33.687,33.375)(-.03158,-.089928){10}{\line(0,-1){.089928}}
\multiput(33.371,32.476)(-.030499,-.101416){9}{\line(0,-1){.101416}}
\multiput(33.096,31.563)(-.03323,-.132043){7}{\line(0,-1){.132043}}
\multiput(32.864,30.639)(-.031707,-.155657){6}{\line(0,-1){.155657}}
\multiput(32.673,29.705)(-.029495,-.188329){5}{\line(0,-1){.188329}}
\put(32.526,28.763){\line(0,-1){.9474}}
\put(32.421,27.816){\line(0,-1){.9512}}
\put(32.36,26.865){\line(0,-1){2.8563}}
\put(32.438,24.008){\line(0,-1){.9465}}
\multiput(32.55,23.062)(.031082,-.188073){5}{\line(0,-1){.188073}}
\multiput(32.705,22.122)(.033018,-.155384){6}{\line(0,-1){.155384}}
\multiput(32.903,21.189)(.030049,-.115289){8}{\line(0,-1){.115289}}
\multiput(33.144,20.267)(.031353,-.101155){9}{\line(0,-1){.101155}}
\multiput(33.426,19.357)(.032338,-.089659){10}{\line(0,-1){.089659}}
\multiput(33.749,18.46)(.033082,-.080083){11}{\line(0,-1){.080083}}
\multiput(34.113,17.579)(.033639,-.071951){12}{\line(0,-1){.071951}}
\multiput(34.517,16.716)(.0316145,-.0602945){14}{\line(0,-1){.0602945}}
\multiput(34.96,15.872)(.032041,-.0548716){15}{\line(0,-1){.0548716}}
\multiput(35.44,15.049)(.0323517,-.0500196){16}{\line(0,-1){.0500196}}
\multiput(35.958,14.248)(.0325626,-.0456407){17}{\line(0,-1){.0456407}}
\multiput(36.511,13.472)(.0326862,-.0416587){18}{\line(0,-1){.0416587}}
\multiput(37.1,12.722)(.0327324,-.0380138){19}{\line(0,-1){.0380138}}
\multiput(37.722,12)(.0327093,-.0346584){20}{\line(0,-1){.0346584}}
\multiput(38.376,11.307)(.0342549,-.0331316){20}{\line(1,0){.0342549}}
\multiput(39.061,10.644)(.0376098,-.0331958){19}{\line(1,0){.0376098}}
\multiput(39.776,10.014)(.0412549,-.0331943){18}{\line(1,0){.0412549}}
\multiput(40.518,9.416)(.0452381,-.0331196){17}{\line(1,0){.0452381}}
\multiput(41.287,8.853)(.0496194,-.0329623){16}{\line(1,0){.0496194}}
\multiput(42.081,8.326)(.0544748,-.0327111){15}{\line(1,0){.0544748}}
\multiput(42.898,7.835)(.0599025,-.0323511){14}{\line(1,0){.0599025}}
\multiput(43.737,7.382)(.0660312,-.0318634){13}{\line(1,0){.0660312}}
\multiput(44.595,6.968)(.073033,-.031223){12}{\line(1,0){.073033}}
\multiput(45.472,6.593)(.089256,-.033434){10}{\line(1,0){.089256}}
\multiput(46.364,6.259)(.100763,-.03259){9}{\line(1,0){.100763}}
\multiput(47.271,5.966)(.114912,-.03146){8}{\line(1,0){.114912}}
\multiput(48.19,5.714)(.13283,-.029931){7}{\line(1,0){.13283}}
\multiput(49.12,5.504)(.187678,-.033385){5}{\line(1,0){.187678}}
\multiput(50.059,5.337)(.23626,-.031){4}{\line(1,0){.23626}}
\put(51.004,5.214){\line(1,0){.9497}}
\put(51.953,5.133){\line(1,0){.9524}}
\put(52.906,5.095){\line(1,0){.9531}}
\put(53.859,5.101){\line(1,0){.9518}}
\put(54.811,5.151){\line(1,0){.9486}}
\multiput(55.759,5.244)(.188676,.027185){5}{\line(1,0){.188676}}
\multiput(56.703,5.38)(.156034,.029796){6}{\line(1,0){.156034}}
\multiput(57.639,5.558)(.132441,.031609){7}{\line(1,0){.132441}}
\multiput(58.566,5.78)(.114505,.03291){8}{\line(1,0){.114505}}
\multiput(59.482,6.043)(.090309,.030476){10}{\line(1,0){.090309}}
\multiput(60.385,6.348)(.080751,.031418){11}{\line(1,0){.080751}}
\multiput(61.273,6.693)(.072632,.032144){12}{\line(1,0){.072632}}
\multiput(62.145,7.079)(.0656229,.0326959){13}{\line(1,0){.0656229}}
\multiput(62.998,7.504)(.0594886,.0331061){14}{\line(1,0){.0594886}}
\multiput(63.831,7.967)(.0540567,.0333974){15}{\line(1,0){.0540567}}
\multiput(64.642,8.468)(.0491985,.0335872){16}{\line(1,0){.0491985}}
\multiput(65.429,9.006)(.0448157,.033689){17}{\line(1,0){.0448157}}
\multiput(66.191,9.579)(.0408318,.0337134){18}{\line(1,0){.0408318}}
\multiput(66.926,10.185)(.037187,.0336688){19}{\line(1,0){.037187}}
\multiput(67.632,10.825)(.0338332,.0335622){20}{\line(1,0){.0338332}}
\multiput(68.309,11.496)(.0322684,.0350693){20}{\line(0,1){.0350693}}
\multiput(68.954,12.198)(.032249,.0384247){19}{\line(0,1){.0384247}}
\multiput(69.567,12.928)(.0321567,.0420687){18}{\line(0,1){.0420687}}
\multiput(70.146,13.685)(.0319828,.0460488){17}{\line(0,1){.0460488}}
\multiput(70.689,14.468)(.0317166,.0504248){16}{\line(0,1){.0504248}}
\multiput(71.197,15.275)(.0335834,.0592204){14}{\line(0,1){.0592204}}
\multiput(71.667,16.104)(.0332225,.0653579){13}{\line(0,1){.0653579}}
\multiput(72.099,16.953)(.032727,.072371){12}{\line(0,1){.072371}}
\multiput(72.492,17.822)(.032067,.080495){11}{\line(0,1){.080495}}
\multiput(72.844,18.707)(.031201,.090061){10}{\line(0,1){.090061}}
\multiput(73.156,19.608)(.030071,.101543){9}{\line(0,1){.101543}}
\multiput(73.427,20.522)(.032673,.132182){7}{\line(0,1){.132182}}
\multiput(73.656,21.447)(.03105,.155789){6}{\line(0,1){.155789}}
\multiput(73.842,22.382)(.028701,.188451){5}{\line(0,1){.188451}}
\put(73.986,23.324){\line(0,1){.9478}}
\put(74.086,24.272){\line(0,1){1.7282}}
\end{picture}
\caption{\footnotesize
The spectrum of the operator $H$. The spectrum out of the large disc
is real.}
\lb{Figspec}
\end{figure}
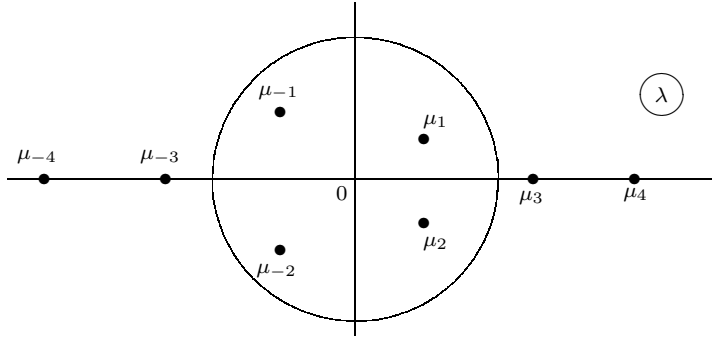
In the unperturbed case $p=q=0$ all eigenvalues $\m_n$ are simple, real and
have the form
\[
\lb{unpev}
\m_{n}^o=(\n n)^3,\qq n\in\Z\sm\{0\},\qqq \n={2\pi\/\sqrt3}.
\]

For the function $f\in L^1(\T)$ we introduce the Fourier coefficients
\[
\lb{wtFc}
\wt f_n=
{2\/\sqrt3}\int_0^1 f(x)\cos \Big(2\pi nx+{\pi\/6}\Big)dx,\qq n\in\N.
\]
We formulate our first main results about asymptotics of the
eigenvalues.

\begin{theorem}
\lb{Thevas}
Let $p,q\in L^1(\T)$.
Then each eigenvalue $\m_n $ with $|n|$ large enough is
real and has algebraic multiplicity one. Moreover,
\[
\lb{asmn1}
\m_n=\m_{n}^o-2\n n p_0+\n n\wt p_n+O(n^{1\/2}),
\]
as $n\to\pm\iy$.
If, in addition, $p'\in L^1(\T)$, then
the eigenvalues $\m_n$ satisfy
\[
\lb{asmnss0}
\m_n=\m_{n}^o-2\n np_0 +\n n\wt p_n+q_0-\wt q_n +O(n^{-{1\/2}}).
\]
Moreover, if $p'',q'\in L^1(\T)$, then
\[
\lb{asmn0}
\m_n=\m_{n}^o-2\n np_0 +\n n\wt p_n+q_0-\wt q_n
+{4p_0^2\/3\n n}+O(n^{-{3\/2}}).
\]
\end{theorem}

In the nice paper \cite{McK81} McKean studied inverse problems
for third order operators on the circle. In particular,
he determined the trace formula
for the operator $H$ in the case $p,q\in C^\iy(\T)$.
We extend this formula to a larger class of coefficients.
Consider the shifted operator
$
H_t=H(p_t,q_t),
$
given in \er{Hpq},
where $t\in\T$,
$$
p_t(x)=p(x+t),\qq q_t(x)=q(x+t)\qq\forall\ x\in\R.
$$
The spectrum consists of eigenvalues
$\m_n(t),n\in\Z\sm\{0\},...\le \Re\m_{-2}(t)\le \Re\m_{-1}(t)
\le \Re\m_{1}(t)\le \Re\m_{2}(t)\le...$
counting with multiplicities.

\begin{theorem}
\lb{ThTrf}
Let $p''',q''\in L^1(\T)$. Then there exists $N=N(p,q)\in\N$ such that the functions
$\sum_{n=1}^N\m_{n}(t)$ and each
$\m_n(t),n>N$, belong to the space $ C^1(\T)$. Moreover, the following
trace formula holds true:
\[
\lb{trftr}
\sum_{n=-\iy}^\iy\big(\m_n(t)-\m_n(0)\big)=V(0)-V(t),
\]
the series converges absolutely and uniformly in $t\in\T$, where
\[
\lb{deffh}
V=q-{p'\/3}.
\]
In particular, assume that we know $\m_{n}(t)$
for all $(n,t)\in \Z\ts\T$.
Then

a) If, in addition, we know $p$ and $q(0)$, then we can recover $q$.

b) If, in addition, we know $q,p'(0)$ and $p(0)$, then we can recover $p$.
\end{theorem}

\no {\bf Remark.}
1) McKean \cite{McK81} obtained the trace formula in the case $p,q\in C^\iy(\T)$,
however, he does not discuss convergence of the series.

2) The proof of Theorem uses the methods from our paper \cite{BK15}.

3) Due to Lemma \ref{CLev}~ii), each eigenvalue $\m_n(t)$ with $|n|$
large enough is simple and it is a smooth function of $t\in\T$.
Moreover, it is shown in \cite{McK81} that in the case small $p,q\in
C^\iy(\T)$ the motion of the large eigenvalues is quite similar to the
motion of the Dirichlet eigenvalues for the Schr\"odinger operators
$-y''+qy$, see Trubowitz~\cite{T77} for the potential $q\in C^3(\T)$
and Korotyaev~\cite{K99} for the potential $q\in L^2(\T)$. The
situation for the eigenvalues inside the bounded disc is more
complicated, see the example with the $\d$-coefficients in
Section~\ref{Sectdelta}. However, due to Rouch\'e's theorem, we can
control their sum and it is smooth.

3) Let $D(\l,t)=D(\l,p_t,q_t)$ and let $n\in\Z\sm\{0\}$.
The identity $D(\m_n(t),t)=0$ gives that
each function $\m_n(t)$ satisfies the so-called Dubrovin equation
for the operator $H$
$$
\dot\m_n(t)=-{\dot D(\l,t)\/D'(\l,t)}\Big|_{\l=\m_n(t)},
\qqq t\in\T,
$$
where $\dot D={\pa D\/\pa t},D'={\pa D\/\pa\l}$. Dynamics of the
eigenvalues plays an important role in solving the Boussinesq equation.
It will be discussed in a separate paper.

\medskip

The plan of the paper is as follows. The proof of Theorem~\ref{Thevas}
is rather complicated and in Section~\ref{Sect2} we give
a sketch of proof of this Theorem.
Section~\ref{Sect3} contains some preliminary simplest relations
for the characteristic function $D$ and for the fundamental matrix $M$.
Moreover, the unperturbed case $p=q=0$ and the example with $p=0$
and $q$ is the periodic $\d$-function
are considered there.
Section~\ref{SectBM} is devoted to the Birkhoff method,
which is a main tool of our proofs of the eigenvalue asymptotics.
We present this method in a general formulation, since in the next parts of the paper
it will be applied in three different cases, depending on the smoothness of the coefficients.
In Section~\ref{Sectev} we consider the eigenvalues for the case $p,q\in L^1(\T)$.
There we prove the Counting result and determine the asymptotics for this case.
The eigenvalue asymptotics for the case $p',q\in L^1(\T)$
is determined in Section~\ref{Sect6} and for the case $p'',q'\in L^1(\T)$
in Section~\ref{Sect7}.
We prove the trace formula in Section~\ref{Sect8}.

\section{Sketch of proofs}
\setcounter{equation}{0}
\lb{Sect2}

\subsection{Factorization formula}
In this Section we describe briefly our proof of Theorem~\ref{Thevas}.
Our main tool is an asymptotic analysis of the fundamental matrix $M$
at large $|\l|$.
Such analysis is standard for Schr\"odinger
operators.  For this case (even with the matrix
coefficients) all  entries of the fundamental matrix are bounded for
$\l\to+\iy$ (in the unperturbed case they have the form
$\cos \sqrt\l x,\sin \sqrt\l x$). But in our case we meet additional
difficulties. For
our third order operator all entries of the fundamental
matrix are unbounded as $\l\to+\iy$
(in the unperturbed case they have the form \er{fsunp}).
In order to obtain the asymptotics of the fundamental matrix we use the method
developed by Birkhoff \cite{B08}.
Now we give a brief description
of this method, see the details in Section~\ref{SectBM}.

Let $z=\l^{1\/3}\in\cZ_+$, where
\[
\lb{defZ+}
\cZ_+=\Big\{z\in\C:\arg z\in\big[0,{\pi\/3}\big)\Big\}.
\]
Introduce the diagonal matrix $\cT$ and the matrix $\Omega$ by
\[
\lb{4g.Om}
\cT=\diag(\t_1,\t_2,\t_3)=\diag(\t,\t^2,1),\qqq \t=e^{i{2\pi\/3}},
\]
\[
\lb{defmaZ}
\Omega=\ma1&1&1\\\t z&\t^2 z&z \\
\t^2z^2&\t z^2&z^2\am.
\]
The key point in our proof of the asymptotics is the following
factorization formula \er{4g.rmmipr00} for the matrix-valued solution
of equation  \er{me1}.

\begin{theorem}
\lb{Thfact0}
Let $p,q\in L^1(\T)$. Then there exists a matrix-valued
solution $A(x,z)$ of equation  \er{me1} such that each function
$A(x,\cdot),x\in [0,2]$ is analytic in $\cZ_+$ for $|z|$ large enough and
satisfies
\[
\lb{4g.rmmipr00}
A(x,\l)=\O(z)\big(\1_3+O(z^{-1})\big)e^{zx\cT},
\]
as $|z|\to\iy,z\in\cZ_+$, uniformly in $(\arg z,
x)\in[0,{\pi\/3}]\ts [0,2]$.

\end{theorem}

The factorization formula \er{4g.rmmipr00} represents the
matrix-valued solution
of equation  \er{me1} in the form of product of the bounded matrices
and the diagonal matrix, which contains all exponentially increasing
and exponentially decreasing factors.

\subsection{Sketch of proof of asymptotics \er{4g.rmmipr00}}
Let $z\in\cZ_+$ and
let $A(x,z),x\in\R,$ be a matrix-valued solution of equation \er{me1}.
We show how to determine the solution $A$ with the needed
asymptotics \er{4g.rmmipr00} using three steps:

{\it Step 1.}
We introduce the matrix-valued function
$Y(x,z)$ by
\[
\lb{defcM}
A(x,z)= \Omega(z)Y(x,z).
\]
Substituting the definition \er{defcM} into equation \er{me1}
and using the identity
$$
\Omega^{-1}\cP\Omega
=z\cT+{\cQ\/z},\qqq \cQ=-{p\/3}\ma \t^2&-\t&-1\\
-\t^2&\t&-1\\
-\t^2&-\t&1\am-{q\/3z}\ma\t&\t&\t\\\t^2&\t^2&\t^2\\ 1& 1& 1\am,
$$
we obtain that $Y$ satisfies the equation
\[
\lb{eqcMi}
Y'-z\cT Y={\cQ\/z} Y.
\]
It is easy to find the inverse operator for the operator
on the left side of equation \er{eqcMi}.
This would be sufficient in the case of a second order operator.
However, for higher order operators, additional steps are required.

{\it Step 2.} We introduce a matrix-valued function $ X(x,z)$
by
\[
\lb{cAcXi}
 Y(x,z)= X(x,z) e^{zx\cT}.
\]
Then $ X$ is a solution of
the differential equation
\[
\lb{4g.ecGipri}
 X'+z( X\cT-\cT  X)={\cQ\/z} X.
\]
Equation \er{4g.ecGipri}, as well as the equivalent equations
\er{me1} and \er{eqcMi}, has many solutions. In order to find the
unique one we choose the solution under the two-side initial
conditions
$$
 X_{jk}(0,z)=0,\qq j<k,\qqq  X_{jk}(2,z)=\d_{jk},\qq j\ge k.
$$
This choice of solution $X$ will lead us to the desired solution $A$ of equation \er{me1}
satisfying the asymptotics \er{4g.rmmipr00} and this is a crucial point of the method.

{\it Step 3.} Let $z\in\cZ_+$ and let $|z|$ be large enough.
It is shown in Theorem~\ref{Thf} that $ X$ is the unique solution of the integral equation
\[
\lb{ineqX}
 X=\1_{3}+{1\/z}K X,
\]
where
$$
(K X)_{\ell j}(x,z)=\int_0^2K_{\ell j}(x,s,z)
(\cQ  X)_{\ell j}(s,z)ds
\qqq\forall\qq\ell ,j=1,2,3,
$$
$$
K_{\ell j}(x,s,z)= \ca \ \
e^{z(x-s)(\t_\ell-\t_j)}\chi(x-s),
\ \ \  \ell <j\\
-e^{z(x-s)(\t_\ell-\t_j)}\chi(s-x), \ \ \
\ell \ge j\ac,\qq \chi(s)=\ca 1,\ s\ge 0\\ 0,\ s<0\ac.
$$
Note that
\[
\lb{4g.esom}
\Re(\t_1z)\le\Re(\t_2z)\le\Re(\t_3z)\qq \forall\ \ z\in\cZ_+.
\]
Then the kernel of the integral operator $K$ satisfies
$|K_{\ell j}(x,s,z)|\le1$ for all
$\ell,j=1,2,3$ and $(x,s,z)\in[0,2]^2\ts\cZ_+$.

Iterations of the integral equation \er{ineqX} give the asymptotics
\[
\lb{asX}
X(x,z)=\1_3+O(z^{-1})
\]
as $|z|\to\iy,z\in\cZ_+$ uniformly in $x\in[0,2]$.
Then the definition \er{cAcXi} gives
\[
\lb{asY}
 Y(x,z)=\big(\1_3+O(z^{-1})\big) e^{zx\cT}.
\]
Substituting this asymptotics into \er{defcM} we obtain
the asymptotics \er{4g.rmmipr00}.

\subsection{Asymptotics of the characteristic function}
Let $(x,z)\in[0,2]\ts\cZ_+$ and let $|z|$ be large enough.
Then the fundamental matrix $M(x,\l)$
satisfies the identity
\[
\lb{MsimPhii}
M(x,\l)= A(x,z) A^{-1}(0,z),\qqq \l=z^3.
\]
The functions
\[
\lb{fsphii}
\phi_j(x,z)= A_{1j}(x,z),\qq j=1,2,3,
\]
are solutions of equation \er{1b}. In Lemma~\ref{Lmrel}
we will prove that
\[
\lb{idsigmai}
D(\l)={\det\phi(z)\/\det A(0,z)},
\]
where
\[
\lb{defphizi}
\phi(z)=\ma
\phi_1(0,z)&\phi_2(0,z)&\phi_3(0,z)\\
\phi_1(1,z)&\phi_2(1,z)&\phi_3(1,z)\\
\phi_1(2,z)&\phi_2(2,z)&\phi_3(2,z)
\am.
\]
The asymptotics \er{4g.rmmipr00} (see Lemma~\ref{CLev}~i)) gives
\[
\lb{asdetFii}
\det A(0,z)=-i3\sqrt3z^3 \big(1+O(z^{-1})\big),
\]
as $|z|\to\iy,z\in\cZ_+$. This asymptotics and the identity
\er{idsigmai} show that the large positive zeros of the function $D$
coincide with the zeros of the third order determinant
$\det\phi(z)$.

Moreover,
substituting the asymptotics \er{4g.rmmipr00} into the definition
\er{fsphii} and using \er{defmaZ} we obtain
\[
\lb{asphijepri}
\phi_j(x,z)=e^{zx\t_j}\big(1+O(z^{-1})\big),
\]
as $|z|\to\iy,z\in\cZ_+$ uniformly in $x\in[0,2]$.
These asymptotics show that the solutions $\phi_j$ are Jost type
solutions of equation \er{1b}.

For simplicity, consider the real $z\to+\iy$, see Lemma~\ref{LmasA}
for the situation in the whole sector $\cZ_+$.
The definition \er{4g.Om} gives that $\phi_1,\phi_2,\phi_3$
satisfy
$$
|\phi_1(x,z)|=|e^{zx\t}|\big(1+O(z^{-1})\big)
=e^{-{1\/2}zx}\big(1+O(z^{-1})\big),
$$
$$
|\phi_2(x,z)|=|e^{zx\t^2}|\big(1+O(z^{-1})\big)
=e^{-{1\/2}zx}\big(1+O(z^{-1})\big),
$$
and
$$
|\phi_3(x,z)|=e^{xz}\big(1+O(z^{-1})\big).
$$
Substituting these asymptotics into
the definition \er{defphizi} we obtain
$$
\det\phi(z)=\phi_3(2,z)\Big(\det\ma
\phi_1(0,z)&\phi_2(0,z)\\
\phi_1(1,z)&\phi_2(1,z)
\am+O(e^{-z})\Big).
$$
Substituting the last asymptotics and the asymptotics
\er{asdetFii} into the identity \er{idsigmai} we get
the following asymptotic reprsentation for the function $D$:
\[
\lb{asDlam}
D(\l)={i\phi_3(2,z)\/3\sqrt3z^3}\Big(\det\ma
\phi_1(0,z)&\phi_2(0,z)\\
\phi_1(1,z)&\phi_2(1,z)
\am+O(e^{-z})\Big),
\]
as $z\to+\iy$.
Thus, we reduce the asymptotic analysis of the zeros
of the third-order determinant $\det\phi(z)$
in the identity \er{idsigmai} to the analysis of the zeros of
the simpler second-order determinant in \er{asDlam}.

Of course, the rough asymptotics \er{asphijepri} does not give us
asymptotics of the zeros of the function $D$. However,
we can take the next terms in the iteration series for the solution
of the integral equation \er{ineqX}. It improves the asymptotics of
the matrix-valued function $A$ and then the asymptotics of
$\phi_1$ and $\phi_2$, due to the definition \er{fsphii}.
In this way, using the identity \er{asDlam} we determine the asymptotics
\er{asmn1}--\er{asmn0} of the zeros of the function $D$.

\section{Characteristic function $D$}
\setcounter{equation}{0}
\lb{Sect3}

\subsection{Properties of the characteristic function}
In the following Lemma we establish some simple properties of the function $D$.
Moreover, there we prove for completeness that the spectrum of the operator
$H$ is the set of zeros
of the entire function $D(\l)$ given by the definition \er{defsi}.

Denote by $D(\l,p,q)$ the function $D(\l)$ for the operator $H=H(p,q)$
given by \er{Hpq}--\er{cDH} and let  $\m_n(p,q)$ be zeros of the function $D(\l,p,q)$.

\begin{lemma}
\lb{Lmspec}
Let $p,q\in L^1(\T)$.
Then

i) The spectrum $\s(H)$ of the operator $H$ satisfies
\er{spec}.

ii)  The function $D(\l)$ is real for $\l\in\R$ and satisfies
\[
\lb{ssymia}
D(\l,p,q)=D(-\l,p^-,-q^-)\qqq \forall\ \ \l\in\C,
\]
where $p^-(t)=p(-t),q^-(t)=q(-t),t\in\R$.
The zeros $\m_n(p,q)$ of the function $D(\l,p,q)$
satisfy
\[
\lb{symev}
\m_{-n}(p,q)=-\m_n(p^-,-q^-)\qqq\forall\ \ n\in\Z\sm\{0\}.
\]

\end{lemma}

\no {\bf Proof.}
i) Assume that $y(x,\l),(x,\l)\in[0,2]\ts\C$
is the solution of equation \er{1b}
satisfying the three-point conditions
\[
\lb{mpvcond}
y(0,\l)=y(1,\l)=y(2,\l)=0.
\]
Then
$$
y(x,\l)=C_1\vp_1(x,\l)+C_2\vp_2(x,\l)+C_3\vp_3(x,\l),
$$
where $\vp_j(x,\l)$ are the fundamental solutions and $C_1,C_2,C_3$ are complex constant.
The conditions \er{mpvcond} yield $C_1=0$ and
$$
C_2\vp_2(1,\l)+C_3\vp_3(1,\l)=0,\qq C_2\vp_2(2,\l)+C_3\vp_3(2,\l)=0.
$$
The solution $y(x,\l)$ is non-trivial iff $D(\l)=0$.
The function $D$ is entire, then the spectrum is pure discrete and satisfies
\er{spec}.

ii) The definition \er{defsi} shows that $D(\l)$ is real for $\l\in\R$.
Substituting $-t$ instead of $t$ in equation \er{1b}
we obtain that the operator $H(p,q)$ is unitarily equivalent
to the operator $-H(p^-,-q^-)$.
This yields \er{ssymia}. The identity \er{symev} follows.
\BBox

\medskip

\no {\bf Remark.}
Using these results we reduce the analysis of the function
$D$ in $\C$ to the analysis in the domain $\Re\l\ge 0,\Im\l\ge 0$
and the analysis of the eigenvalues $\m_n$ at $n\to-\iy$
to the analysis at $n\to+\iy$.

\subsection{The unperturbed case}
\lb{SSunpc}
If $p=q=0$, then the fundamental solutions have the forms
\[
\lb{fsunp}
\begin{aligned}
\vp_1^o(x,\l)={1\/3}(e^{zx}+ e^{\t zx}+e^{\t^2 zx}),\\
\vp_2^o(x,\l)={1\/3z}(e^{zx}+\t^2 e^{\t zx}+\t e^{\t^2 zx}),\\
\vp_3^o(x,\l)={1\/3z^2}(e^{zx}+\t e^{\t zx}+\t^2 e^{\t^2 zx})
\end{aligned}
\]
here and below
$$
\t=e^{i{2\pi\/3}},\qqq z=\l^{1\/3},\qqq z\in\ol\cZ,
$$
$$
\begin{aligned}
\ol\cZ=\Big\{z\in \C:\arg z\in\Big(-{\pi\/3},{\pi\/3}\Big]\Big\},\\
\cZ=\Big\{z\in \C:\arg z\in\Big(-{\pi\/3},{\pi\/3}\Big)\Big\}.
\end{aligned}
$$
Then for the function $D$ at $p=q=0$ we have
\[
\lb{idsig0}
\begin{aligned}
D^o={1\/9\l}
\det\ma e^{z}+\t^2 e^{\t z}+\t e^{\t^2 z}&e^{z}+\t e^{\t z}+\t^2 e^{\t^2 z}\\
e^{2z}+\t^2 e^{2\t z}+\t e^{2\t^2 z}&e^{2z}+\t e^{2\t z}+\t^2 e^{2\t^2 z}\am
\\
={\t-\t^2\/9\l}
\det\ma 1&1&1\\e^{z}& e^{\t z}&e^{\t^2 z}\\
e^{2z}& e^{2\t z}&e^{2\t^2 z}\am={i\/3\sqrt3\l}
\det\ma 1&1&1\\e^{z}& e^{\t z}&e^{\t^2 z}\\
e^{2z}& e^{2\t z}&e^{2\t^2 z}\am.
\end{aligned}
\]
The standard formula for the Vandermonde determinant gives
$$
D^o={i\/3\sqrt3\l}(e^{z}-e^{\t z})(e^{z}-e^{\t^2 z})(e^{\t
z}-e^{\t^2 z}) =-{8\/3\sqrt3 \l}\sin{\sqrt 3 z\/2}\sin{\sqrt 3 \t
z\/2} \sin{\sqrt 3 \t^2z\/2}.
$$
Here only the first factor $\sin{\sqrt 3 \/2}z$ has zeros in $\ol\cZ$ and then
the zeros $\m_{n}^o, n\in\Z\sm\{0\}$, of $D^o$ are simple and
have the form \er{unpev}.

\subsection{Example: $\d$-coefficient}
\lb{Sectdelta}

\begin{figure}
\centering
\tiny
\unitlength 0.8mm
\linethickness{0.2pt}
\ifx\plotpoint\undefined\newsavebox{\plotpoint}\fi 
\begin{picture}(177.5,86.5)(10,20)
\put(6.25,54){\line(1,0){171.25}}
\put(87.75,86.5){\line(0,-1){59.25}}
\put(22.25,52.5){\line(0,1){2.0}}
\put(152.25,53.5){\line(0,1){2.0}}
\put(142.5,50.5){\makebox(0,0)[cc]{$({2\pi\/\sqrt3})^3$}}
\put(33,57.){\makebox(0,0)[cc]{$-({2\pi\/\sqrt3})^3$}}
\put(86,50.25){\makebox(0,0)[cc]{$0$}}
\put(96,55.25){\makebox(0,0)[cc]{$a_1$}}
\put(126,52.25){\makebox(0,0)[cc]{$a_2$}}
\put(91.,78.){\makebox(0,0)[cc]{$\displaystyle +$}}
\put(95.5,78.){\makebox(0,0)[cc]{$r_{-1}^+$}}
\put(91.,30.){\makebox(0,0)[cc]{$\displaystyle +$}}
\put(95.,30.){\makebox(0,0)[cc]{$r_0^-$}}
\put(111.,68.){\makebox(0,0)[cc]{$\displaystyle +$}}
\put(115.,68.){\makebox(0,0)[cc]{$r_0^+$}}
\put(111.,40.){\makebox(0,0)[cc]{$\displaystyle +$}}
\put(115.5,40.){\makebox(0,0)[cc]{$r_1^-$}}
\put(22.25,56.5){\makebox(0,0)[cc]{$r_{-1}^-$}}
\put(152.25,51.25){\makebox(0,0)[cc]{$r_1^+$}}
\put(167.5,83.75){\makebox(0,0)[cc]{$\l$}}
\put(167.5,83.75){\circle{5.315}}
\linethickness{0.5pt}
\put(32.75,53.0){\circle*{2.828}}
\put(32.75,52.5){\line(1,0){65.50}}
\put(75.,52.5){\vector(1,0){.07}}
\put(101.25,35.75){\vector(1,-2){.07}}\qbezier(98.25,52.5)(98.25,29.625)(110.25,29.75)
\put(119.75,35.75){\vector(1,2){.07}}\qbezier(122.75,53.5)(122.75,29.625)(110.75,29.75)
\put(122.75,53.5){\line(-1,0){100.30}}
\put(75.,53.5){\vector(-1,0){.07}}
\put(25.,53.5){\vector(-1,0){.07}}
\put(22.25,52.5){\line(1,0){30.0}}
\put(30,52.5){\vector(1,0){.07}}
\put(52,47.25){\makebox(0,0)[cc]{$\m_{-1}(t)$}}
\put(142.75,55.0){\circle*{2.828}}
\put(142.75,54.5){\line(-1,0){44.50}}
\put(130.,54.5){\vector(-1,0){.07}}
\put(101.25,72.25){\vector(1,2){.07}}\qbezier(98.25,54.5)(98.25,78.375)(110.25,78.25)
\put(119.75,72.25){\vector(1,-2){.07}}\qbezier(122.75,55.5)(122.75,78.375)(110.75,78.25)
\put(122.75,55.5){\line(1,0){29.50}}
\put(132.,55.5){\vector(1,0){.07}}
\put(151.,55.5){\vector(1,0){.07}}
\put(152.25,54.5){\line(-1,0){9.0}}
\put(147.,54.5){\vector(-1,0){.07}}
\put(142,61.25){\makebox(0,0)[cc]{$\m_1(t)$}}
\end{picture}
\caption{\footnotesize
The motion of the eigenvalues $\m_{\pm1}(t)$  of the operator
with the $\d$-coefficient as $t$ runs the interval $[0,1]$, $\g=40$.
Black circles are the positions of the eigenvalues at $t=0$ and $t=1$,
the points $r_0^\pm,r_{\pm1}^\pm$ are branch points of the multiplier curve.}
\lb{Figevm}
\end{figure}
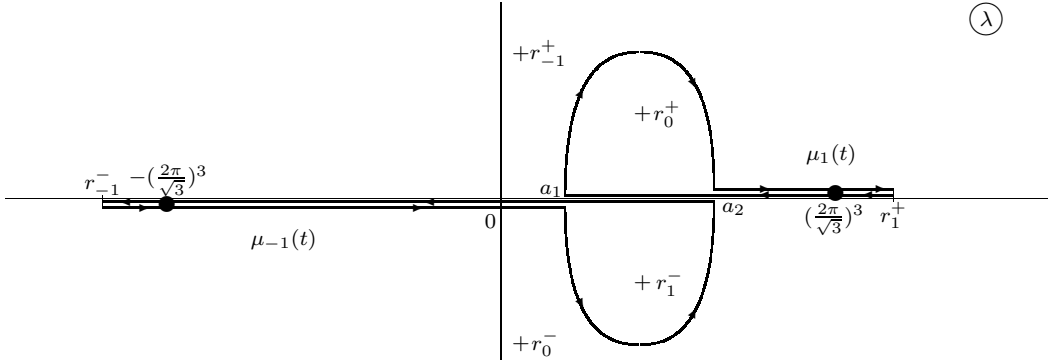

In order to illustrate the movement of eigenvalues we consider an example.
Consider the operator $H_t$ with a periodic $\d$-potential,
where $p=0,q=\gamma\sum_{n=-\infty}^{\infty}\delta(x+t-n)$.
The standard calculations show that the fundamental solutions $\varphi_2,\varphi_3$ satisfy
$$
\begin{aligned}
&\varphi_j(1,\l,t)=\varphi_j^o(1,\l)-\gamma\varphi^o_j(t,\l)\varphi_3^o(1-t,\l),
\\
&\varphi_j(2,\l,t)=\varphi_j^o(2,\l)-\gamma\varphi^o_j(t,\l)\varphi_3^o(2-t,\l)
-\gamma\varphi^o_3(1-t,\l)\big(\varphi^o_j(t+1,\l)
-\gamma\varphi_j^o(t,\l)\varphi_3^o(1,\l)\big).
\end{aligned}
$$
The results of numeric analysis of the zeros $\m_{\pm 1}(t)$ of the function
$D(\l,t)$, defined by \er{defsi}, are shown in Fig.~\ref{Figevm}.
Here we take $\g=40$ and have
$$
\begin{aligned}
\m_{-1}(0)=\m_{-1}(1)=-\Big({2\pi\/\sqrt3}\Big)^3,\qqq
\m_1(0)=\m_1(1)=\Big({2\pi\/\sqrt3}\Big)^3,
\\
\m_{-1}(t_1)=\m_1(t_1),\qq \m_{-1}(t_2)=\m_1(t_2),
\end{aligned}
$$
for some $0<t_1<t_2<1$ and
$$
\m_{-1}(t)=\ol\m_1(t)\qq\forall\ t\in(t_1,t_2).
$$
Moreover, the extreme positions of $\m_{\pm 1}(t)$ coincide
with the branch points $r_{\pm 2}$ of the so called multiplier curve
for the operator with the periodic coefficients on the axis, see \cite{McK81}.
Thus, the eigenvalue $\m_1(t)$:

1) starts as $t=0$ from the point $({2\pi\/\sqrt3})^3$
moving to the left,

2) collides with the eigenvalue $\m_{- 1}(t)$ at the point
$a_1$ as $t=t_1$,

3) leaves the real axis and moves along the oval curve in the complex plane,

4) returning to the real axis it collides with the eigenvalue $\m_{- 1}(t)$ again
at the point $a_2$ as $t=t_2$,

5) moves along the real axis to the right until the branch point $r_2$,

5) turns back and returns to the point $({2\pi\/\sqrt3})^3$ as $t=1$.

\no The motion of $\m_{-1}(t)$ is similar, see  Fig.~\ref{Figevm}.
The other eigenvalues move similarly to the eigenvalues of the Dirichlet
problem for the Scr\"odinger operator associated with
the Korteweg-de Vries equation. Branch points of the multiplier curve
satisfy $...<r_{-2}^-<r_{-2}^+<r_{-1}^-<r_1^+<r_2^-<r_2^+<r_3^-<r_3^+...$
If $t$ changes from $0$ to $1$,
then the eigenvalue $\m_n(t),n=\pm 2, \pm 3,...$
runs all interval $[r_n^-,r_n^+]$ making $n$ complete revolutions.

Thus, the movement of the first eigenvalues is complicated.
It can start moving along the real axis, then go out into the complex plane,
and then return to the real axis again.
When the parameter $\g$ becomes larger, the movement
of the eigenvalues becomes even more complicated.
Therefore, the analysis of such eigenvalues is difficult.

\subsection{Fundamental solutions}
The matrix-valued function $M(\l)$ is entire, however it is difficult
to obtain its asymptotics at large $|\l|$.
We introduce other matrix-valued solutions of equation \er{me1}
that differ from the solution $M$.
We take a $3\ts 3$ matrix-valued function $A(x,z),z\in \cZ$, such that

1) $A(\cdot,z)$
satisfies equation \er{me1} on $[0,2]$,

2) each $A(x,\cdot),x\in[0,2]$
is analytic on a domain $\mD\ss \cZ$ and

3) $\det A(0,z)\ne 0$ for all $z\in \mD$.

\no We can always achieve the fulfillment of the condition 3) by choosing
the domain $\mD$ that is  sufficiently small.

The fundamental matrix $M$ satisfies the identity
\[
\lb{MsimPhi}
M(x,\l)= A(x,z) A^{-1}(0,z),\qqq\forall\qq (x,z)\in[0,2]\ts \mD.
\]
More often we need the entries
from the first line
of the matrix-valued function $A$.
The functions
\[
\lb{fsphi}
\phi_j(x,z)= A_{1j}(x,z),\qq j=1,2,3,
\]
are fundamental solutions of equation \er{1b}.
Introduce the matrix-valued function $\phi$ by
\[
\lb{defphiz}
\phi(z)=\ma
\phi_1(0,z)&\phi_2(0,z)&\phi_3(0,z)\\
\phi_1(1,z)&\phi_2(1,z)&\phi_3(1,z)\\
\phi_1(2,z)&\phi_2(2,z)&\phi_3(2,z)
\am.
\]
The functions $\det\phi(z)$ and $\det A(0,z)$ are analytic in $\mD$,
but, in general, they are not entire.

\begin{lemma}
\lb{Lmrel}
Let $p,q\in L^1(\T)$.
Then the function $D$ has the form
\[
\lb{idsigma}
D(\l)={\det\phi(z)\/\det A(0,z)},\qqq z=\l^{1\/3}\in\mD,
\]
and the function ${\displaystyle{\det\phi(z)\/\det A(0,z)}}$
has an analytic extension from $\mD$ onto the whole sector $\cZ$.
Thus, the identity \er{idsigma} holds for all $z\in\cZ,\l=z^3\in\C$.
\end{lemma}

\no {\bf Proof.}  The function $ A$ is a solution of
equation \er{me1}, then
\[
\lb{fr1}
 A=\ma\phi_1&\phi_2&\phi_3\\
\phi_1'&\phi_2'&\phi_3'\\
\e_1&\e_2&\e_3\am,\qq\e_j=\phi_j''+p\phi_j.
\]
For each $j=1,2,3$ we have
$$
\phi_j(x,z)=\phi_j(0,z)\vp_1(x,z)+\phi_j'(0,z)\vp_2(x,z)
+\e_j(0,z)\vp_3(x,z).
$$
Then
\[
\lb{fr2}
\ma
\phi_1(0,z)&\phi_2(0,z)&\phi_3(0,z)\\
\phi_1(1,z)&\phi_2(1,z)&\phi_3(1,z)\\
\phi_1(2,z)&\phi_2(2,z)&\phi_3(2,z)
\am
=\ma
\vp_1(0,\l)&\vp_2(0,\l)&\vp_3(0,\l)\\
\vp_1(1,\l)&\vp_2(1,\l)&\vp_3(1,\l)\\
\vp_1(2,\l)&\vp_2(2,\l)&\vp_3(2,\l)
\am\ma\phi_1&\phi_2&\phi_3\\
\phi_1'&\phi_2'&\phi_3'\\
\e_1&\e_2&\e_3\am(0,z).
\]
The definition \er{defsi} implies the identity
\[
\lb{fr3}
D(\l)=\det\ma
1&0&0\\
\vp_1(1,\l)&\vp_2(1,\l)&\vp_3(1,\l)\\
\vp_1(2,\l)&\vp_2(2,\l)&\vp_3(2,\l),
\am=\det\ma
\vp_1(0,\l)&\vp_2(0,\l)&\vp_3(0,\l)\\
\vp_1(1,\l)&\vp_2(1,\l)&\vp_3(1,\l)\\
\vp_1(2,\l)&\vp_2(2,\l)&\vp_3(2,\l)
\am.
\]
Substituting the identities \er{fr1} and \er{fr3} into \er{fr2}
we obtain \er{idsigma}.
\BBox

\medskip

\no{\bf Remark.} The asymptotics \er{asdetFii} shows that $\det A(0,z)$ does not
vanish at large $|z|$ and then
the zeros of the function $D(z^3)$ coincide with the zeros
of $\det\phi(z)$ at high energy.

\section{Birkhoff's method}
\setcounter{equation}{0}
\lb{SectBM}

\subsection{Birkhoff's differential equation}
In Section~\ref{Sect2} we rewrote the problem \er{me1} in the form \er{eqcMi}, i.e.,
\[
\lb{eqcM}
Y'-z\cT Y=-{1\/3z}\Big(pP+{q\/z} Q\Big) Y,
\qqq Y(0,z)=\1_3,
\]
where
\[
\lb{cK12}
P=\ma \t^2&-\t&-1\\
-\t^2&\t&-1\\
-\t^2&-\t&1\am,
\qq
 Q=\ma\t&\t&\t\\\t^2&\t^2&\t^2\\ 1& 1& 1\am.
\]
We analyze equation \er{eqcM} by using our version of the Birkhoff
method of asymptotic analysis of higher-order equations,
see \cite{BK11}, \cite{BK14x}.
Now we present this method for third order case.

Introduce the domains
\[
\lb{defZ+r}
\cZ_+(r)=\{z\in \cZ_+:|z|>r\}\ss \cZ_+,\qqq r>0,
\]
where the sector $\cZ_+$ has the form \er{defZ+}.

\begin{condition}
\lb{DefcQTh}
Let a $3\ts 3$~-~matrix-valued functions $\Phi(x,z)$
and $\Theta(x,z),(x,z)\in\T\ts \cZ_+$, satisfy:

1) $\Theta=\diag(\Theta_1,\Theta_2,\Theta_3)$ is diagonal,

2) $\Phi(\cdot,z)\in L^1(\T)$
and $\Theta(\cdot,z)\in L^1(\T)$ for all  $z\in \cZ_+(r)$, where
$r>0$ is large enough,

3) $\Phi(x,z)$ and $\Theta(x,z)$ are analytic in
$\cZ_+(r)$ and
\[
\lb{ascQTh}
\Phi(x,z)=\cF(x)+O(z^{-1}),\qqq \Theta(x,z)=\cT+O(z^{-1})
\]
as $|z|\to\iy,z\in \cZ_+$,
uniformly in $x$, where $\cT$
is given by \er{4g.Om} and $\cF\in L^1(\T)$,

4) $\cF$ is off-diagonal, that is $\cF_{jj}=0,j=1,2,3$.
\end{condition}

We consider the following differential
equation on $\R$:
\[
\lb{me2pr}
\cA'-z\Theta\cA={1\/z^m}\Phi\cA,
\]
where $z\in \cZ_+(r)$,
$r>0$ is large enough, $m\in\N$, and the $3\ts 3$-matrix-valued functions
$\Theta(x,z)$ and $\Phi(x,z)$
satisfy Conditions~\ref{DefcQTh}.
Below $\Theta(x,z)$, $\Phi(x,z)$ and $m$ have different forms depending on smoothness of
the coefficients $p,q$. For example, if $p,q\in L^1(\T)$, then equation \er{me2pr}
has the form \er{eqcM}.

\medskip

\no {\bf Remark.} 1) Here and always below our asymptotics at large $|z|$ are uniform
in $\arg z$.

2) The hypothesis 4 in Condition~\ref{DefcQTh}
is assumed without loss of generality,
since we can replace the diagonal part of $\cF$ into the left hand side
of equation \er{me2pr}.

\medskip

Birkhoff \cite{B08}
developed a method for obtaining asymptotics of
fundamental solutions of higher-order linear differential equations.
He found a way to rewrite the differential equation \er{me1}
in the form of a Fredholm integral equation with a small kernel at high energy.
Birkhoff used a scalar form of the differential equation.
However, the matrix form \er{me2pr} is more convenient for analysis.
Below we give our version of the Birkhoff method for
third order differential equations,
the case of fourth order equations see in \cite{BK14x}.

First of all, we rewrite equation \er{me2pr} in the form \er{4g.ecGipr} convenient
for the application of the Birkhoff method.

\begin{lemma}
Let the $3\ts 3$~-~matrix-valued functions $\Phi$
and $\Theta$ satisfy Condition~\ref{DefcQTh} and let
$z\in \cZ_+(r)$, where $r>0$ is large enough.
Let a matrix-valued function $\cA$ and a matrix-valued function $\cX$
satisfy
\[
\lb{cAcX}
\cA (x,z)=\cX(x,z)e^{z\int_0^x\Theta(s,z)ds},\qqq x\in\R.
\]
Then $\cA$ is a solution of the differential equation \er{me2pr} if and only if
$\cX$ is a solution of
the differential equation
\[
\lb{4g.ecGipr}
\cX'+z(\cX\Theta-\Theta \cX)={1\/z^m}\Phi \cX
\]
\end{lemma}

\no {\bf Proof.}
 Let the matrix-valued function $\cA$
satisfy the differential equation  \er{me2pr} and let
\er{cAcX} be fulfilled. Then
$$
(\cX e^{z\int_0^x\Theta ds})'-z\Theta \cX e^{z\int_0^x\Theta ds}
={1\/z^m}\Phi \cX e^{z\int_0^x\Theta ds},
$$
which gives
$$
\cX' e^{z\int_0^x\Theta ds}+z\cX \Theta e^{z\int_0^x\Theta ds}
-z\Theta \cX e^{z\int_0^x\Theta ds}
={1\/z^m}\Phi \cX e^{z\int_0^x\Theta ds}.
$$
Thus, $\cX$ satisfies the differential equation \er{4g.ecGipr}.

Conversely, let the matrix-valued function $\cA$ have the form \er{cAcX}
and let $\cX$ satisfy the differential equation \er{4g.ecGipr}.
The identities
$$
\begin{aligned}
\cA'=z\cX\Theta e^{z\int_0^x\Theta ds}
+\cX'e^{z\int_0^x\Theta ds}
=z\cX \Theta e^{z\int_0^x\Theta ds}
-\Big(z(\cX\Theta-\Theta \cX)-{1\/z^m}\Phi \cX\Big)e^{z\int_0^x\Theta ds}
\\
=\Big(z\Theta \cX+{1\/z^m}\Phi \cX\Big)e^{z\int_0^x\Theta ds}
=z\Theta \cA+{1\/z^m}\Phi \cA
\end{aligned}
$$
show that $\cA$ satisfies the differential equation \er{me2pr}.
\BBox

\subsection{Birkhoff's integral equation}
\lb{SectBm}
Below in Theorem~\ref{Thf}~i) we will prove that the differential equation \er{4g.ecGipr}
is equivalent in some sense to an integral equation.
But first we will study this integral equation.
We introduce the following class of the Birkhoff operators.

\begin{definition}
\lb{DetK}

Let $K$ be an integral operator in
the space $C[0,2]$ of $3\ts 3$ continuous matrix-valued functions
on $[0,2]$ given by
\[
\lb{4g.dcLipr}
(K\cX)_{\ell j}(x,z)=\int_0^2K_{\ell j}(x,s,z)
(\Phi \cX)_{\ell j}(s,z)ds
\qqq\forall\qq\ell ,j=1,2,3,
\]
where $\Phi$ and $\Theta$ satisfy Condition~\ref{DefcQTh} and
\[
\lb{Kljpr} K_{\ell j}(x,s,z)= \ca \ \
e^{z\int_s^x(\Theta_\ell(u,z)-\Theta_j(u,z))du}\chi(x-s),
\ \ \  \ell <j\\
-e^{z\int_s^x(\Theta_\ell(u,z)-\Theta_j(u,z))du}\chi(s-x), \ \ \
\ell \ge j\ac,\qq \chi(s)=\ca 1,\ s\ge 0\\ 0,\ s<0\ac.
\]
\end{definition}

If $|z|$ is large, then, due to \er{4g.esom},
the kernel of the integral operator $K$ is bounded.
Therefore, the matrix-valued integral equation
\[
\lb{4g.me5ipr}
\cX=\1_{3}+{1\/z^m}K\cX,\qqq m\in\N,
\]
that we call the Birkhoff equation,
has a unique solution for large $|z|$.

\begin{lemma}
\lb{4g.lmwtG0}
Let the $3\ts 3$~-~matrix-valued functions $\Phi$
and $\Theta$ satisfy the Condition~\ref{DefcQTh}
and let $K$ be the integral operator given by Definition~\ref{DetK}.

i)
Let $z\in \cZ_+(r)$, where $r>0$ is large enough, and let $m\in\N$. Then
the integral equation \er{4g.me5ipr}
has the unique solution $\cX(\cdot,z)\in C[0,2]$.
Moreover, each matrix-valued function
$\cX(x,\cdot),x\in[0,2]$, is analytic in $\cZ_+(r)$ and satisfies
\[
\lb{4g.aswtGpr}
\cX(x,z)=\1_{3}+{\cB(x,z)\/z^m}+{O(1)\/z^{2m}},
\]
as $|z|\to\iy,z\in \cZ_+$, uniformly on $x\in[0,2]$, where
\[
\lb{4g.itGipr}
\cB=K\1_3.
\]

ii) The function $\cB=K\1_3$ satisfies
\[
\lb{asGjjpr}
\cB_{j j}(x,z)=O(z^{-1}),\qqq j=1,2,3,
\]
\[
\lb{asGljpr}
\begin{aligned}
\cB_{\ell j}(x,z)
=-\int_x^2e^{-z(s-x)(\t_\ell-\t_j)}
\cF_{\ell j}(s)ds+O(z^{-1}),
\qqq 1\le j<\ell\le 3,
\\
\cB_{\ell j}(x,z)
=\int_0^xe^{z(x-s)(\t_\ell-\t_j)}
\cF_{\ell j}(s)ds+O(z^{-1}),
\qqq 1\le \ell<j\le 3,
\end{aligned}
\]
as $|z|\to\iy,z\in \cZ_+$ uniformly in $x\in[0,2]$.
Moreover, the matrix-valued functions
\[
\lb{defcGj}
\eta_j(x,z)=\sum_{k=1}^3\cB_{kj}(x,z),\qq j=1,2,
\]
satisfy
\[
\lb{idG1f}
\begin{aligned}
\eta_1(x,z)=\int_0^xe^{i2\pi n(x-s)}\cF_{12}(s)ds+O(n^{-{1\/2}}),\\
\eta_2(x,z)=-\int_x^2e^{i2\pi n(s-x)}\cF_{21}(s)ds+O(n^{-{1\/2}}),
\end{aligned}
\]
as $z={2\pi n\/\sqrt3}+O(n^{-1}),n\to+\iy$ uniformly in $x\in[0,2]$.

\end{lemma}

\no {\bf Proof.} i) Let $z\in \cZ_+(r)$, where $r>0$ is large enough.
Due to the asymptotics \er{ascQTh} and the estimates \er{4g.esom},
the function $K(x,s,z)$ satisfies $|K(x,s,z)|\le C$
for all $(x,s,z)\in [0,2]^2\ts\cZ_+(r)$ and for some $C>0$. Then
the operator $K$ is bounded in $C[0,2]$.
Then the operator ${1\/z^m}K$ in equation \er{4g.me5ipr} is a contraction.
Consider the iteration series
\[
\lb{4g.me6ipr}
\cX(x,z)=\sum_{n=0}^\iy
{(K^n\1_3)(x,z)\/z^{nm}}\qqq\forall\qq(x,z)\in[0,2]\ts \cZ_+(r).
\]
The matrix-valued functions $K^n\1_3,n\ge 0$, satisfy
$\|K^n\1_3\|_{C[0,2]}\le C$ for some $C>0$, where $\|\cdot\|_{C[0,2]}$
is any matrix norm.
Then the series \er{4g.me6ipr} converges absolutely and uniformly on any
compact set in $[0,2]\ts \cZ_+(r)$.
Each matrix valued function
$K^n\1_3(x,\cdot),n\ge 0,t\in[0,2]$, is analytic in $\cZ_+(r)$.
Then each matrix valued function
$\cX(x,\cdot),x\in[0,2]$, is analytic in $\cZ_+(r)$ and satisfies
the asymptotics \er{4g.aswtGpr}.

ii) The definitions \er{4g.dcLipr} and \er{4g.itGipr} and the asymptotics \er{ascQTh}
 give
$$
\cB_{\ell j}(x,z)
=\int_0^2K_{\ell j}(x,s,z)\cF_{\ell j}(s)ds+O(z^{-1}),\qqq
\ell,j=1,2,3.
$$
The definition \er{Kljpr} implies \er{asGjjpr} and \er{asGljpr}.

Let $z={2\pi n\/\sqrt3}+O(n^{-1}),n\to+\iy$.
Then the asymptotics \er{asGljpr} yield
\[
\lb{asGlj12}
\begin{aligned}
\cB_{12}(x,z)
=\int_0^xe^{i2\pi n(x-s)}\cF_{12}(s)ds+O(n^{-1}),
\\
\cB_{21}(x,z)
=-\int_x^2e^{i2\pi n(s-x)}\cF_{21}(s)ds+O(n^{-1}),
\end{aligned}
\]
and
\[
\lb{asg131pr}
\cB_{3 1}(x,z)
=-\int_x^2e^{-(\sqrt3-i)\pi n(s-x)}
\cF_{31}(s)ds+O(n^{-1}).
\]
Moreover,
$$
\begin{aligned}
\Big|\int_x^2e^{-(\sqrt3-i)\pi n(s-x)}\cF_{31}(s)ds\Big|
\le\int_x^2e^{-\sqrt3\pi n(s-x)}|\cF_{31}(s)|ds
\\
\le\Big(\int_x^2e^{-2\sqrt3\pi n(s-x)}ds\Big)^{1\/2}
\Big(\int_x^2|\cF_{31}(s)|^2ds\Big)^{1\/2}=O(n^{-{1\/2}})
\end{aligned}
$$
uniformly in $x$.
Then the asymptotics \er{asg131pr} implies
\[
\lb{asGlj3}
\cB_{3 1}(x,z)=O(n^{-{1\/2}}),
\]
uniformly in $x\in[0,2]$.
Similarly,
\[
\lb{asGlj3a}
\cB_{3 2}(x,z)=O(n^{-{1\/2}}).
\]
The asymptotics \er{asGjjpr},
\er{asGlj12}, \er{asGlj3} and \er{asGlj3a} yield
the asymptotics \er{idG1f}.
\BBox

\subsection{Factorization theorem}
Let $z\in \cZ_+(r)$, where $r>0$ is large enough.
Introduce the fundamental matrix-valued solution $\cM$ of equation \er{me2pr}
by the condition
$$
\cM(0,z)=\1_3.
$$
Then
\[
\lb{cYcAcA}
\cM(x,z)=\cA(x,z)\cA^{-1}(0,z).
\]
Now we prove the basic result of this Section,
the factorization formula \er{4g.rmmipr} for the solution
$\cM$. In fact, this formula together with
the asymptotics \er{4g.aswtGpr} gives the high energy asymptotics
of the fundamental matrix, see Remark below.

\begin{theorem}
\lb{Thf}
Let the $3\ts 3$~-~matrix-valued functions $\Phi$
and $\Theta$ satisfy Condition~\ref{DefcQTh} and let
$z\in \cZ_+(r)$, where $r>0$ is large enough.

i) The matrix-valued function $\cX(x,z),x\in[0,2]$, satisfies the integral equation
\er{4g.me5ipr} if and only if it satisfies the differential equation \er{4g.ecGipr}
and the initial conditions
\[
\lb{inccX}
\cX_{jk}(0,z)=0,\qq j<k,\qqq \cX_{jk}(2,z)=\d_{jk},\qq j\ge k.
\]

ii) Let a matrix-valued function
$\cX(x,z)$
and a matrix-valued function $\cM(x,z)$ satisfy
\[
\lb{4g.rmmipr}
\cM(x,z)=\cX(x,z)e^{z\int_0^x\Theta(s,z)ds}\cX^{-1}(0,z),\qqq \forall\  x\in[0,2].
\]
Then $\cM (x,z)$ is the solution of the initial problem $\cM (0,z)=\1_3$ for equation
\er{me2pr} if and only if
$\cX(x,z)$ is the solution
of the integral equation \er{4g.me5ipr}.
\end{theorem}

\no {\bf Proof.}
i) Let $\cX$ satisfy equation \er{4g.ecGipr}.
In terms of matrix entries we have
\[
\lb{4g.ecGipr1}
\cX_{jk}'+z(\cX_{jk}\Theta_k-\Theta_j \cX_{jk})
={1\/z^m}\sum_{\ell=1}^3\Phi_{j\ell} \cX_{\ell k},\qqq j,k=1,2,3.
\]
Assume, in addition, that $\cX$ satisfies the initial conditions \er{inccX}.
Integrating equation \er{4g.ecGipr1} and using the conditions \er{inccX}
we obtain
\[
\lb{iecXjk}
\begin{aligned}
\cX_{jk}(x)={1\/z^m}\int_0^xe^{z\int_s^x(\Theta_k(u)-\Theta_j(u))du}
\sum_{\ell=1}^3\Phi_{j\ell}(s) \cX_{\ell k}(s)ds,\qqq j<k
\\
\cX_{jk}(x)=-{1\/z^m}\int_x^2e^{z\int_s^x(\Theta_k(u)-\Theta_j(u))du}
\sum_{\ell=1}^3\Phi_{j\ell}(s) \cX_{\ell k}(s)ds,\qqq j>k
\end{aligned}
\]
for all $j,k=1,2,3$, and
\[
\lb{iecXjj}
\cX_{jj}(x)=1-{1\/z^m}\int_x^2
\sum_{\ell=1}^3\Phi_{j\ell}(s) \cX_{\ell j}(s)ds,\qq j=1,2,3.
\]
Using the definitions \er{Kljpr} we rewrite these equations in the form
\[
\lb{iecXjk0}
\cX_{jk}=\d_{jk}+{1\/z^m}\int_0^2K_{jk}(x,s,z)
(\Phi \cX)_{jk}(s,z)ds,\qqq j,k=1,2,3.
\]
The definition  \er{4g.dcLipr} yields that
the function $\cX$ satisfies equation \er{4g.me5ipr}.

Conversely, let $\cX$ satisfy equation \er{4g.me5ipr}.
In terms of matrix entries equation \er{4g.me5ipr} has the form \er{iecXjk0}.
Substituting the definitions \er{Kljpr}
into  \er{iecXjk0} we obtain that the functions $\cX_{jk}$
satisfy \er{iecXjk}, \er{iecXjj},
which yields \er{inccX}.
Differentiating the identities \er{iecXjk}, \er{iecXjj}, we get that $\cX_{jk}$
satisfy equations \er{4g.ecGipr1}.
Then $\cX$ satisfies equation \er{4g.ecGipr}.

ii) Let the matrix-valued function $\cX$ be the solution
of the integral equation \er{4g.me5ipr}. Then $\cX$ satisfies
the differential equation \er{4g.ecGipr}
and the initial conditions \er{inccX}. Then the matrix $\cM$, given by \er{cYcAcA},
satisfies the differential equation \er{me2pr} and the initial condition
$\cM(0,z)=\1_3$.

Conversely, let $\cM$ satisfy equation \er{me2pr} and let $\cM$ has the the form
 \er{4g.rmmipr}.
Substituting \er{4g.rmmipr} into \er{me2pr} we obtain that $\cX$
satisfies the differential equation \er{4g.ecGipr},
and then $\cX(x,z)$ is the solution
of the integral equation \er{4g.me5ipr}.
\BBox

\medskip

\no {\bf Remark.} Recall that
the matrix-valued functions $\cX(x,z)$ and $\cX^{-1}(x,z)$ are uniformly bounded
on $[0,2]\ts\cZ_+(r)$, see Lemma~\ref{4g.lmwtG0}~i). Thus
the formula \er{4g.rmmipr} represents the fundamental matrix
in the form of product of the bounded matrices
$\cX(x,z)$ and $\cX^{-1}(0,z)$  and the diagonal matrix $e^{z\int_0^x\Theta(s,z)ds}$
which contains all exponentially increasing and exponentially decreasing
factors. We have the asymptotics \er{4g.aswtGpr} of the function $\cX$,
then we have asymptotics of the fundamental matrix $M$.

\subsection{Asymptotics of the solutions $\phi_j$}
\lb{Sectassol}
Always in this paper, see \er{idPhi0}, \er{idPhiss} and \er{idPhi},
the matrix-valued function
$ A$ in \er{MsimPhi} has the form
\[
\lb{idPhiGen}
 A(x,z)=\Omega(z) \big(\1_3+{1\/z^2}\cW(x,z)\big)\cX(x,z)
e^{z\int_0^x\Theta(s,z)ds},\qq (x,z)\in [0,2]\ts \cZ_+,
\]
where $\cW=0$ in \er{idPhi0},
$\cW={p\/3}W_1$ in \er{idPhiss} and $\cW={p\/3}W_1+{1\/3z}W_2$ in \er{idPhi}.
In the following Lemma we determine asymptotics of
the Jost type solutions $\phi_j$, given by \er{fsphi},
when $ A$ satisfies \er{idPhiGen}.

\begin{lemma}
\lb{LmAsphi}
Let $p,q\in L^1(\T)$.
Let the $3\ts 3$ matrix-valued function $\cW(x,z)$ satisfies:
$\cW(\cdot,z)\in L^1(\T)$ for all $z\in \cZ_+$, $\cW(x,\cdot)$ is analytic
in $\cZ_+$ and $\cW(x,z)=O(1)$ as $|z|\to\iy,z\in \cZ_+$ uniformly in $x\in\T$.
Let the matrix-valued function  $\cX$ be the solution of equation \er{4g.me5ipr}
for some  $m=1,2,3$ and $\Theta$ and $\Phi$ satisfying Condition~\ref{DefcQTh}.
Let the matrix-valued function $ A$ satisfies \er{idPhiGen},
and let $\phi_j= A_{1j},j=1,2,3$.
Then
\[
\lb{asphijepr}
\phi_j(x,z)=e^{z\int_0^x\Theta_j(s,z)ds}
\big(1+\zeta_{j}(x,z)+O(z^{-m-2})\big),
\]
as $|z|\to\iy,z\in \cZ_+$, uniformly in $x\in[0,2]$,
where
\[
\lb{phi1jpr}
\zeta_{j}(x,z)={\eta_j(x,z)\/z^m}+{1\/z^2}\sum_{k =1}^3\cW_{kj}(x,z)
,
\]
and $\eta_j$ are given by \er{defcGj}.
\end{lemma}

\no {\bf Proof.}
Let $|z|\to\iy,z\in \cZ_+$, and let $m=1,2,3$.
Substituting the definition \er{defmaZ} and
the asymptotics \er{4g.aswtGpr} into
the definition \er{idPhiGen}
we obtain
$$
\begin{aligned}
\phi_j(x,z)= A_{1j}(x,z)
=e^{z\int_0^x\Theta_j(s,z)ds}\sum_{k,\ell =1}^3
\big(\d_{k\ell}+{1\/z^2}\cW_{k\ell}(x,z)\big)\big(\d_{\ell
j}+{1\/z^m}\cB_{\ell j}(x,z)+O(z^{-2m})\big)
\\
=e^{z\int_0^x\Theta_j(s,z)ds}\sum_{k,\ell =1}^3
\Big(\d_{k\ell}\d_{\ell j}+ {1\/z^2}\cW_{k\ell}(x,z)\d_{\ell j}
+{1\/z^m}\d_{k\ell}\cB_{\ell j}(x,z)+O(z^{-2m})\Big),
\end{aligned}
$$
uniformly in $x\in[0,2]$,
which yields \er{asphijepr}.
\BBox

\section{Eigenvalues for $p,q\in L^1(\T)$}
\setcounter{equation}{0}
\lb{Sectev}

\subsection{Factorization of the fundamental matrix}
\lb{Sectffm0}

In this Section we consider the case $p,q\in L^1(\T)$.
We apply Theorem~\ref{Thf} to equation \er{eqcM}
in order to obtain the factorization of the fundamental matrix.
Preliminary, we have to replace the diagonal part of the matrix $P$
from the right side of \er{eqcM} onto the left one in order to get
the off-diagonal matrix $\cF$ in the asymptotics of $\Phi$ in \er{ascQTh}
(see \er{vnsc}, where the matrix $P-2\cT^2$ is off-diagonal).
Then we obtain the following corollary of Theorem~\ref{Thf}
on the factorization of the fundamental matrix.

\begin{corollary} Let $p,q\in L^1(\T)$. Let $r>0$ be large enough and let
$z\in \cZ_+(r)$.
Then the following representation of the fundamental solution holds true:
\[
\lb{factM}
M(x,\l)= A(x,z) A^{-1}(0,z),
\]
for all $x\in [0,2]$, where $\l=z^3$,
\[
\lb{idPhi0}
 A(x,z)=\Omega(z) \cX(x,z) e^{z\int_0^xT_1(s,z)ds},
\]
$\Omega$ is given by the definition \er{defmaZ},
the diagonal $3\ts 3$ matrix-valued function $T_1$ has the form
\[
\lb{4g.wtOss}
T_1=\cT-{2p\/3z^2}\cT^2,
\]
and $\cX$ is the solution of the integral equation \er{4g.me5ipr}
with $m,\Theta$ and $\Phi$ given by
\[
\lb{vnsc}
m=1,\qq \Theta=T_1,\qq\Phi=-{1\/3}\Big(p(P-2\cT^2)+{q\/z} Q\Big).
\]
Each function $ A(x,\cdot),x\in[0,2]$, is analytic in $\cZ_+(r)$.

\end{corollary}

\no {\bf Proof.} Rewrite
equation \er{eqcM} in the form
\[
\lb{eqcX}
Y'-zT_1Y=-{1\/3z}\Big(p(P-2\cT^2)+{q\/z} Q\Big)Y.
\]
Equation \er{eqcX} has the form \er{me2pr} with $m,\Theta,\Phi$
satisfying \er{vnsc}.
Let $Y$ be the solution of equation \er{eqcX} satisfying the condition
$Y(0)=\1_3$.
Theorem~\ref{Thf} shows that $Y$ satisfies the identity \er{4g.rmmipr}.
Substituting \er{4g.rmmipr} into the definition \er{defcM} we obtain
the representation \er{factM}.
\BBox

\medskip

Theorem~\ref{Thfact0} follows immediately from the previous result.

\medskip

\no {\bf Proof of Theorem~\ref{Thfact0}.}
The asymptotics \er{4g.aswtGpr} gives  \er{asX}.
Substituting the asymptotics \er{asX} into the identity \er{idPhi0}
we obtain \er{4g.rmmipr00}.
\BBox

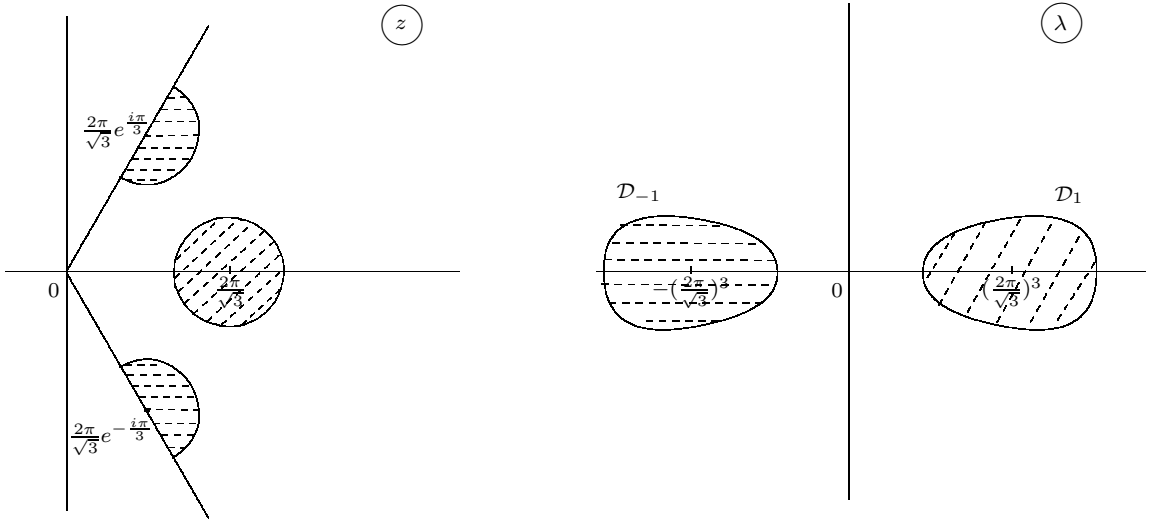
\begin{figure}
\centering
\tiny
\unitlength 0.9mm
\linethickness{0.2pt}
\ifx\plotpoint\undefined\newsavebox{\plotpoint}\fi 
\begin{picture}(178.75,79.425)(10,5)
\put(21.15,82.425){\line(0,-1){72.8}}
\put(12.15,44.825){\line(1,0){66.4}}
\put(52.848,44.765){\line(0,1){.4497}}
\put(52.835,45.215){\line(0,1){.4482}}
\put(52.797,45.663){\line(0,1){.4454}}
\put(52.734,46.108){\line(0,1){.4412}}
\put(52.647,46.55){\line(0,1){.4356}}
\multiput(52.534,46.985)(-.02731,.085721){5}{\line(0,1){.085721}}
\multiput(52.398,47.414)(-.032071,.084056){5}{\line(0,1){.084056}}
\multiput(52.237,47.834)(-.03061,.068438){6}{\line(0,1){.068438}}
\multiput(52.054,48.245)(-.029484,.057099){7}{\line(0,1){.057099}}
\multiput(51.847,48.644)(-.032638,.055356){7}{\line(0,1){.055356}}
\multiput(51.619,49.032)(-.031228,.04676){8}{\line(0,1){.04676}}
\multiput(51.369,49.406)(-.030044,.039943){9}{\line(0,1){.039943}}
\multiput(51.099,49.765)(-.032236,.038197){9}{\line(0,1){.038197}}
\multiput(50.809,50.109)(-.030893,.032697){10}{\line(0,1){.032697}}
\multiput(50.5,50.436)(-.032677,.030914){10}{\line(-1,0){.032677}}
\multiput(50.173,50.745)(-.038176,.03226){9}{\line(-1,0){.038176}}
\multiput(49.829,51.036)(-.039924,.03007){9}{\line(-1,0){.039924}}
\multiput(49.47,51.306)(-.04674,.031258){8}{\line(-1,0){.04674}}
\multiput(49.096,51.556)(-.055336,.032673){7}{\line(-1,0){.055336}}
\multiput(48.709,51.785)(-.05708,.02952){7}{\line(-1,0){.05708}}
\multiput(48.309,51.992)(-.068419,.030654){6}{\line(-1,0){.068419}}
\multiput(47.899,52.176)(-.084035,.032125){5}{\line(-1,0){.084035}}
\multiput(47.479,52.336)(-.085704,.027364){5}{\line(-1,0){.085704}}
\put(47.05,52.473){\line(-1,0){.4355}}
\put(46.614,52.586){\line(-1,0){.4411}}
\put(46.173,52.674){\line(-1,0){.4454}}
\put(45.728,52.737){\line(-1,0){.4482}}
\put(45.28,52.775){\line(-1,0){.8993}}
\put(44.38,52.775){\line(-1,0){.4483}}
\put(43.932,52.738){\line(-1,0){.4455}}
\put(43.487,52.675){\line(-1,0){.4413}}
\put(43.045,52.588){\line(-1,0){.4357}}
\multiput(42.61,52.476)(-.085738,-.027255){5}{\line(-1,0){.085738}}
\multiput(42.181,52.34)(-.084076,-.032018){5}{\line(-1,0){.084076}}
\multiput(41.761,52.179)(-.068458,-.030567){6}{\line(-1,0){.068458}}
\multiput(41.35,51.996)(-.057117,-.029448){7}{\line(-1,0){.057117}}
\multiput(40.95,51.79)(-.055377,-.032603){7}{\line(-1,0){.055377}}
\multiput(40.563,51.562)(-.04678,-.031198){8}{\line(-1,0){.04678}}
\multiput(40.188,51.312)(-.039962,-.030019){9}{\line(-1,0){.039962}}
\multiput(39.829,51.042)(-.038217,-.032211){9}{\line(-1,0){.038217}}
\multiput(39.485,50.752)(-.032717,-.030872){10}{\line(-1,0){.032717}}
\multiput(39.158,50.443)(-.030935,-.032658){10}{\line(0,-1){.032658}}
\multiput(38.848,50.117)(-.032284,-.038156){9}{\line(0,-1){.038156}}
\multiput(38.558,49.773)(-.030095,-.039905){9}{\line(0,-1){.039905}}
\multiput(38.287,49.414)(-.031287,-.04672){8}{\line(0,-1){.04672}}
\multiput(38.036,49.04)(-.032708,-.055315){7}{\line(0,-1){.055315}}
\multiput(37.807,48.653)(-.029557,-.057061){7}{\line(0,-1){.057061}}
\multiput(37.601,48.254)(-.030697,-.068399){6}{\line(0,-1){.068399}}
\multiput(37.416,47.843)(-.032178,-.084015){5}{\line(0,-1){.084015}}
\multiput(37.256,47.423)(-.027419,-.085686){5}{\line(0,-1){.085686}}
\put(37.118,46.995){\line(0,-1){.4354}}
\put(37.006,46.559){\line(0,-1){.4411}}
\put(36.917,46.118){\line(0,-1){.4453}}
\put(36.854,45.673){\line(0,-1){.4482}}
\put(36.815,45.225){\line(0,-1){1.3476}}
\put(36.852,43.877){\line(0,-1){.4455}}
\put(36.914,43.432){\line(0,-1){.4413}}
\put(37.001,42.99){\line(0,-1){.4357}}
\multiput(37.113,42.555)(.027201,-.085756){5}{\line(0,-1){.085756}}
\multiput(37.249,42.126)(.031964,-.084096){5}{\line(0,-1){.084096}}
\multiput(37.409,41.705)(.030523,-.068477){6}{\line(0,-1){.068477}}
\multiput(37.592,41.295)(.029411,-.057136){7}{\line(0,-1){.057136}}
\multiput(37.798,40.895)(.032567,-.055398){7}{\line(0,-1){.055398}}
\multiput(38.026,40.507)(.031168,-.0468){8}{\line(0,-1){.0468}}
\multiput(38.275,40.132)(.029993,-.039982){9}{\line(0,-1){.039982}}
\multiput(38.545,39.773)(.032187,-.038238){9}{\line(0,-1){.038238}}
\multiput(38.835,39.428)(.030852,-.032736){10}{\line(0,-1){.032736}}
\multiput(39.143,39.101)(.032638,-.030956){10}{\line(1,0){.032638}}
\multiput(39.469,38.792)(.038135,-.032309){9}{\line(1,0){.038135}}
\multiput(39.813,38.501)(.039886,-.03012){9}{\line(1,0){.039886}}
\multiput(40.172,38.23)(.0467,-.031317){8}{\line(1,0){.0467}}
\multiput(40.545,37.979)(.055294,-.032743){7}{\line(1,0){.055294}}
\multiput(40.932,37.75)(.057042,-.029593){7}{\line(1,0){.057042}}
\multiput(41.332,37.543)(.06838,-.030741){6}{\line(1,0){.06838}}
\multiput(41.742,37.358)(.083994,-.032232){5}{\line(1,0){.083994}}
\multiput(42.162,37.197)(.085669,-.027474){5}{\line(1,0){.085669}}
\put(42.59,37.06){\line(1,0){.4354}}
\put(43.026,36.947){\line(1,0){.441}}
\put(43.467,36.858){\line(1,0){.4453}}
\put(43.912,36.794){\line(1,0){.4482}}
\put(44.36,36.756){\line(1,0){.8993}}
\put(45.259,36.754){\line(1,0){.4483}}
\put(45.708,36.791){\line(1,0){.4455}}
\put(46.153,36.853){\line(1,0){.4414}}
\put(46.595,36.94){\line(1,0){.4358}}
\multiput(47.03,37.051)(.085773,.027146){5}{\line(1,0){.085773}}
\multiput(47.459,37.187)(.084117,.031911){5}{\line(1,0){.084117}}
\multiput(47.88,37.347)(.068497,.030479){6}{\line(1,0){.068497}}
\multiput(48.291,37.53)(.057155,.029375){7}{\line(1,0){.057155}}
\multiput(48.691,37.735)(.055419,.032532){7}{\line(1,0){.055419}}
\multiput(49.079,37.963)(.04682,.031139){8}{\line(1,0){.04682}}
\multiput(49.453,38.212)(.045001,.033714){8}{\line(1,0){.045001}}
\multiput(49.813,38.482)(.038258,.032163){9}{\line(1,0){.038258}}
\multiput(50.158,38.771)(.032756,.030831){10}{\line(1,0){.032756}}
\multiput(50.485,39.079)(.030976,.032618){10}{\line(0,1){.032618}}
\multiput(50.795,39.406)(.032333,.038114){9}{\line(0,1){.038114}}
\multiput(51.086,39.749)(.030146,.039867){9}{\line(0,1){.039867}}
\multiput(51.357,40.107)(.031347,.04668){8}{\line(0,1){.04668}}
\multiput(51.608,40.481)(.032779,.055273){7}{\line(0,1){.055273}}
\multiput(51.838,40.868)(.029629,.057023){7}{\line(0,1){.057023}}
\multiput(52.045,41.267)(.030784,.06836){6}{\line(0,1){.06836}}
\multiput(52.23,41.677)(.032285,.083974){5}{\line(0,1){.083974}}
\multiput(52.391,42.097)(.027528,.085651){5}{\line(0,1){.085651}}
\put(52.529,42.525){\line(0,1){.4353}}
\put(52.642,42.961){\line(0,1){.441}}
\put(52.731,43.402){\line(0,1){.4453}}
\put(52.795,43.847){\line(0,1){.4481}}
\put(52.834,44.295){\line(0,1){.47}}
\multiput(37.195,46.655)(.0342222,.0326667){18}{\line(1,0){.0342222}}
\multiput(38.427,47.831)(.0342222,.0326667){18}{\line(1,0){.0342222}}
\multiput(39.659,49.007)(.0342222,.0326667){18}{\line(1,0){.0342222}}
\multiput(40.891,50.183)(.0342222,.0326667){18}{\line(1,0){.0342222}}
\multiput(42.123,51.359)(.0342222,.0326667){18}{\line(1,0){.0342222}}
\multiput(36.915,43.855)(.0347368,.0326316){19}{\line(1,0){.0347368}}
\multiput(38.235,45.095)(.0347368,.0326316){19}{\line(1,0){.0347368}}
\multiput(39.555,46.335)(.0347368,.0326316){19}{\line(1,0){.0347368}}
\multiput(40.875,47.575)(.0347368,.0326316){19}{\line(1,0){.0347368}}
\multiput(42.195,48.815)(.0347368,.0326316){19}{\line(1,0){.0347368}}
\multiput(43.515,50.055)(.0347368,.0326316){19}{\line(1,0){.0347368}}
\multiput(44.835,51.295)(.0347368,.0326316){19}{\line(1,0){.0347368}}
\multiput(37.475,41.895)(.0354605,.0331579){19}{\line(1,0){.0354605}}
\multiput(38.822,43.155)(.0354605,.0331579){19}{\line(1,0){.0354605}}
\multiput(40.17,44.415)(.0354605,.0331579){19}{\line(1,0){.0354605}}
\multiput(41.517,45.675)(.0354605,.0331579){19}{\line(1,0){.0354605}}
\multiput(42.865,46.935)(.0354605,.0331579){19}{\line(1,0){.0354605}}
\multiput(44.212,48.195)(.0354605,.0331579){19}{\line(1,0){.0354605}}
\multiput(45.56,49.455)(.0354605,.0331579){19}{\line(1,0){.0354605}}
\multiput(46.907,50.715)(.0354605,.0331579){19}{\line(1,0){.0354605}}
\multiput(38.315,40.075)(.0364087,.0333746){19}{\line(1,0){.0364087}}
\multiput(39.698,41.343)(.0364087,.0333746){19}{\line(1,0){.0364087}}
\multiput(41.082,42.611)(.0364087,.0333746){19}{\line(1,0){.0364087}}
\multiput(42.465,43.879)(.0364087,.0333746){19}{\line(1,0){.0364087}}
\multiput(43.849,45.148)(.0364087,.0333746){19}{\line(1,0){.0364087}}
\multiput(45.232,46.416)(.0364087,.0333746){19}{\line(1,0){.0364087}}
\multiput(46.616,47.684)(.0364087,.0333746){19}{\line(1,0){.0364087}}
\multiput(47.999,48.952)(.0364087,.0333746){19}{\line(1,0){.0364087}}
\multiput(49.383,50.221)(.0364087,.0333746){19}{\line(1,0){.0364087}}
\multiput(39.575,38.535)(.0368421,.0333746){19}{\line(1,0){.0368421}}
\multiput(40.975,39.803)(.0368421,.0333746){19}{\line(1,0){.0368421}}
\multiput(42.375,41.071)(.0368421,.0333746){19}{\line(1,0){.0368421}}
\multiput(43.775,42.339)(.0368421,.0333746){19}{\line(1,0){.0368421}}
\multiput(45.175,43.608)(.0368421,.0333746){19}{\line(1,0){.0368421}}
\multiput(46.575,44.876)(.0368421,.0333746){19}{\line(1,0){.0368421}}
\multiput(47.975,46.144)(.0368421,.0333746){19}{\line(1,0){.0368421}}
\multiput(49.375,47.412)(.0368421,.0333746){19}{\line(1,0){.0368421}}
\multiput(50.775,48.681)(.0368421,.0333746){19}{\line(1,0){.0368421}}
\multiput(41.675,37.555)(.035,.0322){20}{\line(1,0){.035}}
\multiput(43.075,38.843)(.035,.0322){20}{\line(1,0){.035}}
\multiput(44.475,40.131)(.035,.0322){20}{\line(1,0){.035}}
\multiput(45.875,41.419)(.035,.0322){20}{\line(1,0){.035}}
\multiput(47.275,42.707)(.035,.0322){20}{\line(1,0){.035}}
\multiput(48.675,43.995)(.035,.0322){20}{\line(1,0){.035}}
\multiput(50.075,45.283)(.035,.0322){20}{\line(1,0){.035}}
\multiput(51.475,46.571)(.035,.0322){20}{\line(1,0){.035}}
\multiput(43.775,36.995)(.0361111,.0327778){18}{\line(1,0){.0361111}}
\multiput(45.075,38.175)(.0361111,.0327778){18}{\line(1,0){.0361111}}
\multiput(46.375,39.355)(.0361111,.0327778){18}{\line(1,0){.0361111}}
\multiput(47.675,40.535)(.0361111,.0327778){18}{\line(1,0){.0361111}}
\multiput(48.975,41.715)(.0361111,.0327778){18}{\line(1,0){.0361111}}
\multiput(50.275,42.895)(.0361111,.0327778){18}{\line(1,0){.0361111}}
\multiput(51.575,44.075)(.0361111,.0327778){18}{\line(1,0){.0361111}}
\multiput(46.715,36.995)(.0345679,.0319753){18}{\line(1,0){.0345679}}
\multiput(47.959,38.146)(.0345679,.0319753){18}{\line(1,0){.0345679}}
\multiput(49.204,39.297)(.0345679,.0319753){18}{\line(1,0){.0345679}}
\multiput(50.448,40.448)(.0345679,.0319753){18}{\line(1,0){.0345679}}
\multiput(51.692,41.599)(.0345679,.0319753){18}{\line(1,0){.0345679}}
\put(135.425,84.3){\line(0,-1){73}}
\put(98.55,44.825){\line(1,0){80.2}}
\put(166.55,81.425){\circle{6}}
\put(70.175,81.175){\circle{6}}
\multiput(21.125,44.875)(.0337398374,.0587398374){615}{\line(0,1){.0587398374}}
\multiput(21.125,44.625)(.0337398374,-.0587398374){615}{\line(0,-1){.0587398374}}
\qbezier(28.875,58.75)(33.563,56.188)(37.5,59.375)
\qbezier(37.5,59.375)(40.438,61.688)(40.5,66.25)
\qbezier(40.5,66.25)(40.25,69.813)(36.625,72.125)
\qbezier(28.875,30.75)(33.563,33.313)(37.5,30.125)
\qbezier(37.5,30.125)(40.438,27.813)(40.5,23.25)
\qbezier(40.5,23.25)(40.25,19.688)(36.625,17.375)
\put(29.68,58.305){\line(1,0){.9167}}
\put(31.513,58.305){\line(1,0){.9167}}
\put(33.346,58.305){\line(1,0){.9167}}
\put(29.68,31.055){\line(1,0){.9167}}
\put(31.513,31.055){\line(1,0){.9167}}
\put(33.346,31.055){\line(1,0){.9167}}
\put(29.93,59.805){\line(1,0){.9844}}
\put(31.898,59.805){\line(1,0){.9844}}
\put(33.867,59.805){\line(1,0){.9844}}
\put(35.836,59.805){\line(1,0){.9844}}
\put(29.93,29.555){\line(1,0){.9844}}
\put(31.898,29.555){\line(1,0){.9844}}
\put(33.867,29.555){\line(1,0){.9844}}
\put(35.836,29.555){\line(1,0){.9844}}
\put(30.68,61.43){\line(1,0){.9167}}
\put(32.513,61.43){\line(1,0){.9167}}
\put(34.346,61.43){\line(1,0){.9167}}
\put(36.18,61.43){\line(1,0){.9167}}
\put(38.013,61.43){\line(1,0){.9167}}
\put(30.68,27.93){\line(1,0){.9167}}
\put(32.513,27.93){\line(1,0){.9167}}
\put(34.346,27.93){\line(1,0){.9167}}
\put(36.18,27.93){\line(1,0){.9167}}
\put(38.013,27.93){\line(1,0){.9167}}
\put(31.68,63.055){\line(1,0){.8889}}
\put(33.457,63.055){\line(1,0){.8889}}
\put(35.235,63.055){\line(1,0){.8889}}
\put(37.013,63.055){\line(1,0){.8889}}
\put(38.791,63.055){\line(1,0){.8889}}
\put(31.68,26.305){\line(1,0){.8889}}
\put(33.457,26.305){\line(1,0){.8889}}
\put(35.235,26.305){\line(1,0){.8889}}
\put(37.013,26.305){\line(1,0){.8889}}
\put(38.791,26.305){\line(1,0){.8889}}
\put(32.43,64.805){\line(1,0){.9028}}
\put(34.235,64.832){\line(1,0){.9028}}
\put(36.041,64.86){\line(1,0){.9028}}
\put(37.846,64.888){\line(1,0){.9028}}
\put(39.652,64.916){\line(1,0){.9028}}
\put(32.43,24.555){\line(1,0){.9028}}
\put(34.235,24.527){\line(1,0){.9028}}
\put(36.041,24.499){\line(1,0){.9028}}
\put(37.846,24.471){\line(1,0){.9028}}
\put(39.652,24.444){\line(1,0){.9028}}
\put(33.805,66.93){\line(1,0){.875}}
\put(35.555,66.93){\line(1,0){.875}}
\put(37.305,66.93){\line(1,0){.875}}
\put(39.055,66.93){\line(1,0){.875}}
\put(33.805,22.43){\line(1,0){.875}}
\put(35.555,22.43){\line(1,0){.875}}
\put(37.305,22.43){\line(1,0){.875}}
\put(39.055,22.43){\line(1,0){.875}}
\put(34.93,68.68){\line(1,0){.95}}
\put(36.83,68.68){\line(1,0){.95}}
\put(38.73,68.68){\line(1,0){.95}}
\put(34.93,20.68){\line(1,0){.95}}
\put(36.83,20.68){\line(1,0){.95}}
\put(38.73,20.68){\line(1,0){.95}}
\put(36.055,70.43){\line(1,0){.8333}}
\put(37.721,70.513){\line(1,0){.8333}}
\put(36.055,18.93){\line(1,0){.8333}}
\put(37.721,18.846){\line(1,0){.8333}}
\multiput(171.305,46.055)(-.0325758,-.0507576){15}{\line(0,-1){.0507576}}
\multiput(170.327,44.532)(-.0325758,-.0507576){15}{\line(0,-1){.0507576}}
\multiput(169.35,43.009)(-.0325758,-.0507576){15}{\line(0,-1){.0507576}}
\multiput(168.373,41.487)(-.0325758,-.0507576){15}{\line(0,-1){.0507576}}
\multiput(167.396,39.964)(-.0325758,-.0507576){15}{\line(0,-1){.0507576}}
\multiput(166.418,38.441)(-.0325758,-.0507576){15}{\line(0,-1){.0507576}}
\multiput(169.055,49.93)(-.0316667,-.0561111){15}{\line(0,-1){.0561111}}
\multiput(168.105,48.246)(-.0316667,-.0561111){15}{\line(0,-1){.0561111}}
\multiput(167.155,46.563)(-.0316667,-.0561111){15}{\line(0,-1){.0561111}}
\multiput(166.205,44.88)(-.0316667,-.0561111){15}{\line(0,-1){.0561111}}
\multiput(165.255,43.196)(-.0316667,-.0561111){15}{\line(0,-1){.0561111}}
\multiput(164.305,41.513)(-.0316667,-.0561111){15}{\line(0,-1){.0561111}}
\multiput(163.355,39.83)(-.0316667,-.0561111){15}{\line(0,-1){.0561111}}
\multiput(162.405,38.146)(-.0316667,-.0561111){15}{\line(0,-1){.0561111}}
\multiput(164.93,51.68)(-.0330882,-.0603992){14}{\line(0,-1){.0603992}}
\multiput(164.003,49.989)(-.0330882,-.0603992){14}{\line(0,-1){.0603992}}
\multiput(163.077,48.297)(-.0330882,-.0603992){14}{\line(0,-1){.0603992}}
\multiput(162.15,46.606)(-.0330882,-.0603992){14}{\line(0,-1){.0603992}}
\multiput(161.224,44.915)(-.0330882,-.0603992){14}{\line(0,-1){.0603992}}
\multiput(160.297,43.224)(-.0330882,-.0603992){14}{\line(0,-1){.0603992}}
\multiput(159.371,41.533)(-.0330882,-.0603992){14}{\line(0,-1){.0603992}}
\multiput(158.444,39.841)(-.0330882,-.0603992){14}{\line(0,-1){.0603992}}
\multiput(157.518,38.15)(-.0330882,-.0603992){14}{\line(0,-1){.0603992}}
\multiput(160.43,52.43)(-.0322421,-.0570437){14}{\line(0,-1){.0570437}}
\multiput(159.527,50.832)(-.0322421,-.0570437){14}{\line(0,-1){.0570437}}
\multiput(158.624,49.235)(-.0322421,-.0570437){14}{\line(0,-1){.0570437}}
\multiput(157.721,47.638)(-.0322421,-.0570437){14}{\line(0,-1){.0570437}}
\multiput(156.819,46.041)(-.0322421,-.0570437){14}{\line(0,-1){.0570437}}
\multiput(155.916,44.444)(-.0322421,-.0570437){14}{\line(0,-1){.0570437}}
\multiput(155.013,42.846)(-.0322421,-.0570437){14}{\line(0,-1){.0570437}}
\multiput(154.11,41.249)(-.0322421,-.0570437){14}{\line(0,-1){.0570437}}
\multiput(153.207,39.652)(-.0322421,-.0570437){14}{\line(0,-1){.0570437}}
\multiput(155.43,52.18)(-.0327381,-.0589286){14}{\line(0,-1){.0589286}}
\multiput(154.513,50.53)(-.0327381,-.0589286){14}{\line(0,-1){.0589286}}
\multiput(153.596,48.88)(-.0327381,-.0589286){14}{\line(0,-1){.0589286}}
\multiput(152.68,47.23)(-.0327381,-.0589286){14}{\line(0,-1){.0589286}}
\multiput(151.763,45.58)(-.0327381,-.0589286){14}{\line(0,-1){.0589286}}
\multiput(150.846,43.93)(-.0327381,-.0589286){14}{\line(0,-1){.0589286}}
\multiput(149.93,42.28)(-.0327381,-.0589286){14}{\line(0,-1){.0589286}}
\multiput(149.013,40.63)(-.0327381,-.0589286){14}{\line(0,-1){.0589286}}
\put(102.055,51.68){\line(1,0){.9609}}
\put(103.977,51.664){\line(1,0){.9609}}
\put(105.898,51.648){\line(1,0){.9609}}
\put(107.82,51.633){\line(1,0){.9609}}
\put(109.742,51.617){\line(1,0){.9609}}
\put(111.664,51.602){\line(1,0){.9609}}
\put(113.586,51.586){\line(1,0){.9609}}
\put(115.508,51.57){\line(1,0){.9609}}
\put(100.93,49.055){\line(1,0){.9943}}
\put(102.918,49.043){\line(1,0){.9943}}
\put(104.907,49.032){\line(1,0){.9943}}
\put(106.896,49.021){\line(1,0){.9943}}
\put(108.884,49.009){\line(1,0){.9943}}
\put(110.873,48.998){\line(1,0){.9943}}
\put(112.862,48.987){\line(1,0){.9943}}
\put(114.85,48.975){\line(1,0){.9943}}
\put(116.839,48.964){\line(1,0){.9943}}
\put(118.827,48.952){\line(1,0){.9943}}
\put(120.816,48.941){\line(1,0){.9943}}
\put(99.68,46.555){\line(1,0){.995}}
\put(101.67,46.545){\line(1,0){.995}}
\put(103.66,46.535){\line(1,0){.995}}
\put(105.65,46.525){\line(1,0){.995}}
\put(107.64,46.515){\line(1,0){.995}}
\put(109.63,46.505){\line(1,0){.995}}
\put(111.62,46.495){\line(1,0){.995}}
\put(113.61,46.485){\line(1,0){.995}}
\put(115.6,46.475){\line(1,0){.995}}
\put(117.59,46.465){\line(1,0){.995}}
\put(119.58,46.455){\line(1,0){.995}}
\put(121.57,46.445){\line(1,0){.995}}
\put(123.56,46.435){\line(1,0){.995}}
\put(99.305,42.93){\line(1,0){.976}}
\put(101.257,42.92){\line(1,0){.976}}
\put(103.209,42.91){\line(1,0){.976}}
\put(105.16,42.901){\line(1,0){.976}}
\put(107.112,42.891){\line(1,0){.976}}
\put(109.064,42.882){\line(1,0){.976}}
\put(111.016,42.872){\line(1,0){.976}}
\put(112.968,42.862){\line(1,0){.976}}
\put(114.92,42.853){\line(1,0){.976}}
\put(116.872,42.843){\line(1,0){.976}}
\put(118.824,42.834){\line(1,0){.976}}
\put(120.776,42.824){\line(1,0){.976}}
\put(122.728,42.814){\line(1,0){.976}}
\put(100.43,40.18){\line(1,0){.9728}}
\put(102.375,40.18){\line(1,0){.9728}}
\put(104.321,40.18){\line(1,0){.9728}}
\put(106.267,40.18){\line(1,0){.9728}}
\put(108.212,40.18){\line(1,0){.9728}}
\put(110.158,40.18){\line(1,0){.9728}}
\put(112.104,40.18){\line(1,0){.9728}}
\put(114.049,40.18){\line(1,0){.9728}}
\put(115.995,40.18){\line(1,0){.9728}}
\put(117.941,40.18){\line(1,0){.9728}}
\put(119.886,40.18){\line(1,0){.9728}}
\put(121.832,40.18){\line(1,0){.9728}}
\put(105.805,37.555){\line(1,0){.9583}}
\put(107.721,37.555){\line(1,0){.9583}}
\put(109.638,37.555){\line(1,0){.9583}}
\put(111.555,37.555){\line(1,0){.9583}}
\put(113.471,37.555){\line(1,0){.9583}}
\put(115.388,37.555){\line(1,0){.9583}}
\multiput(150.805,50.305)(-.0336538,-.0548077){13}{\line(0,-1){.0548077}}
\multiput(149.93,48.88)(-.0336538,-.0548077){13}{\line(0,-1){.0548077}}
\multiput(149.055,47.455)(-.0336538,-.0548077){13}{\line(0,-1){.0548077}}
\multiput(148.18,46.03)(-.0336538,-.0548077){13}{\line(0,-1){.0548077}}
\multiput(147.305,44.605)(-.0336538,-.0548077){13}{\line(0,-1){.0548077}}
\qbezier(146.25,44.625)(146.125,50.438)(157.5,52.5)
\qbezier(157.5,52.5)(171.813,54.938)(171.625,45)
\qbezier(171.625,45)(171.813,34.313)(158.75,36.625)
\qbezier(158.75,36.625)(146.125,38.813)(146.25,44.625)
\qbezier(125,44.625)(125.125,50.438)(113.75,52.5)
\qbezier(113.75,52.5)(99.438,54.938)(99.625,45)
\qbezier(99.625,45)(99.438,34.313)(112.5,36.625)
\qbezier(112.5,36.625)(125.125,38.813)(125,44.625)
\put(19.25,42.125){\makebox(0,0)[cc]{$0$}}
\put(69.975,81.175){\makebox(0,0)[cc]{$z$}}
\put(166.35,81.425){\makebox(0,0)[cc]{$\l$}}
\put(133.75,42.125){\makebox(0,0)[cc]{$0$}}
\put(167.625,56.125){\makebox(0,0)[cc]{$\cD_1$}}
\put(104.75,56.375){\makebox(0,0)[cc]{$\cD_{-1}$}}
\put(32.5,65.){\line(1,0){0.5}}
\put(28.375,65.625){\makebox(0,0)[cc]{${2\pi\/\sqrt3}e^{i\pi\/3}$}}
\put(32.5,24.25){\line(1,0){0.7}}
\put(27.625,20.5){\makebox(0,0)[cc]{${2\pi\/\sqrt3}e^{-{i\pi\/3}}$}}
\put(45,44.5){\line(0,1){1.0}}
\put(45,41.75){\makebox(0,0)[cc]{${2\pi\/\sqrt3}$}}
\put(159.25,44.5){\line(0,1){1.0}}
\put(159.25,41.75){\makebox(0,0)[cc]{$({2\pi\/\sqrt3})^3$}}
\put(112.375,44.5){\line(0,1){1.0}}
\put(112.375,41.75){\makebox(0,0)[cc]{$-({2\pi\/\sqrt3})^3$}}
\end{picture}
\caption{\footnotesize
The domains $\cD_1$ and $\cD_{-1}$ in the $\l$-plane (on the right)
and their pre-images in the $z$-plane (on the left).}
\lb{4g.ztolambda}
\end{figure}

\subsection{Counting Lemma}
Let $p,q\in L^1(\T)$.
Let the matrix-valued function $ A(x,z)$ be given by \er{idPhi0}.
Now we prove the Counting Lemma for the eigenvalues.
Introduce the domains $\cD_n,n\in\Z$, by
\[
\lb{DomcD}
\begin{aligned}
&\cD_{n}=\Big\{\l\in\C:|z-\n n|
<{\n\/4}\Big\},
\\
&\cD_{-n}=\Big\{\l\in\C:\big|z-e^{ i{\pi\/3}}
\n n\big|
<{\n\/4}\Big\}\cup \big\{\l\in\C:\Big|z-e^{-i{\pi\/3}}
\n n\big|
<{\n\/4}\Big\},
\end{aligned}
\]
where $ n\ge 0,\n={2\pi\/\sqrt3}$, see \er{unpev},
and the contours
$$
C_a(r)=\{\l\in\C:|z-a|=r\},\qq a\in\C,\qq r>0.
$$
Below we need the identity
\[
\lb{detOm}
\det\O=3(\t^2-\t)z^3=-i3\sqrt3z^3,
\]
which follows from the definition \er{defmaZ}.

\begin{lemma}
\lb{CLev}
Let $p,q\in L^1(\T)$. Then

i) The solutions $\phi_j$ of equation \er{1b} and  the function
$\det A(0,z)$
satisfy
\[
\lb{asdetFi}
\det A(0,z)=-i3\sqrt3z^3 \big(1+O(z^{-1})\big),
\]
\[
\lb{asphij}
\phi_j(x,z)=e^{zx\t_j}\big(1+O(z^{-1})\big),
\]
as $j=1,2,3,|z|\to\iy,z\in \cZ$, uniformly in $x\in[0,2]$.
Moreover,
\[
\lb{assig0}
D(\l)=D^o(\l)\big(1+O(z^{-1})\big).
\]
as $|\l|\to\iy$ and $\l\in\C\sm\cup_{n\in\Z}\cD_{n}$.

ii) For each integer $N\ge 0$ large enough the function $D$ has
(counting with multiplicities) $2N+1$ zeros on the domain
$\{|\l|<\n^3(N+{1\/4})^3\}$ and for each $|n|>N$ exactly
one simple zero in the domain $\cD_n$. There are no other zeros.

iii) There exists $N\in\N$ such that each $\m_n,|n|>N$, is real.
\end{lemma}

\no {\bf Proof.}
i) We have $m=1$, $\Theta=\cT_1=\cT+O(z^{-2})$ and $\cW=0$ in Lemma~\ref{LmAsphi}.
Then the asymptotics \er{asphijepr} and \er{phi1jpr} give \er{asphij}.
Then
\[
\lb{asphi}
\phi(z)=\ma
1+O(z^{-1})&1+O(z^{-1})&1+O(z^{-1})\\
e^{\t_1z}(1+O(z^{-1}))&e^{\t_2z}(1+O(z^{-1}))&e^{\t_3z}(1+O(z^{-1}))\\
e^{2\t_1z}(1+O(z^{-1}))&e^{2\t_2z}(1+O(z^{-1}))&e^{2\t_3z}(1+O(z^{-1}))
\am.
\]
Let $|z|\to\iy,z\in \cZ_+,\l\in\C\sm\cup_{n\in\Z}\cD_{n}$. Then the identity \er{idsig0}
gives
$$
\det\phi(z)=-i3\sqrt3z^3D^o(\l)\big(1+O(z^{-1})\big).
$$
Moreover,
$$
\det A(0,z)=\det\Omega(z) \det \cX(0,z)
=-i3\sqrt3z^3\big(1+O(z^{-1})\big),
$$
which yields \er{asdetFi}, here we used $\det\O=-i3\sqrt3z^3$, see \er{detOm}.
Substituting these asymptotics
into the identity \er{idsigma}
we obtain \er{assig0} for $\l\in\C_+$. The identity $D(\ol\l)=\ol D(\l)$
gives \er{assig0} for $\l\in\C_-$.

ii) Let $N\ge 0$ be integer and large enough and let $N'>N$ be another
integer. Let $\l$ belong to the contours $C_0(\n^3(N+{1\/4})^3),
C_0(\n^3(N'+{1\/4})^3)$ and $\pa\cD_n$ for all $
|n|>{N}$. Asymptotics \er{assig0} yields
$$
|D(\l)-D^o(\l)|=|D^o(\l)|\Big|{D(\l)\/D^o(\l)}-1\Big|
=|D^o(\l)|O(|z|^{-1})<|D^o(\l)|
$$
on all contours. Hence, by Rouch\'e's theorem,
$D$ has as many zeros, as $D^o$ in each of the
bounded domains and the remaining unbounded domain.
Since $D^o$ has exactly one simple zero
in each $\cD_n,n\in\Z$,
and since $N'>N$ can be chosen arbitrarily large,
the statement follows.

iii) Let $|n|>N$.
The definition \er{labev} and the statement i) show that the zero
$\m_n$ of the function $D$ satisfies:
$\m_n\in\cD_n$. If $\m_n\not\in\R$, then $\ol\m_n$ is also a zero
of $D$ and $\ol\m_n\in\cD_n$. Then there are two zeros of $D$
in $\cD_n$, which contradicts the statement i).
\BBox

\subsection{Rough eigenvalue asymptotics}
Recall that the eigenvalues of the operator $H$ are zeros of the entire
function $D$ given by the definition \er{defsi}.
The identity \er{idsigma} and the asymptotics \er{asdetFi} show that
the large eigenvalues are zeros of the function $\det\phi(z)$.
The function $\det\phi(z)$ is analytic in $\cZ_+(r)$, where $r>0$ is large enough.
Using the asymptotic behavior of the function $\phi_3(x,z)$ at high energy
we reduce in Lemma~\ref{LmasA} the determinant of the $3\ts 3$-matrix $\phi(z)$ to
the determinant of a $2\ts 2$-matrix.

The asymptotics \er{asphij} and the definitions \er{4g.Om} give
\[
\lb{asmphi3}
|\phi_3(x,z)|=e^{x\Re z}\big(1+O(z^{-1})\big),
\]
as $|z|\to\iy,z\in \cZ_+$.
The asymptotics \er{asmphi3} show that
$|\phi_3(x,z)|>0$ for all $(x,z)\in [0,2]\ts \cZ_+(r)$.
Then the function $\xi (z)$, given by
\[
\lb{definA}
\xi (z)={\det\phi(z)\/\phi_3(2,z)},
\]
is analytic in $\cZ_+(r)$, where $r>0$ is large enough.
Let $\m_n$ be the eigenvalue of the operator $H$. Then
the identity
\er{idsigma} and the asymptotics \er{asdetFi} give
\[
\lb{detAev}
\xi (\m_n^{1\/3})=0
\]
for all $n\in\N$ large enough.
The symmetry \er{symev} shows that it is sufficiently to determine
the asymptotics for the large positive eigenvalues.
Introduce the sector
$$
\cZ_+^+=\{z\in\C:\arg z\in[0,\tf{\pi}{6}]\}.
$$

\begin{lemma}
\lb{LmasA}
Let $p,q\in L^1(\T)$.

i) Let, in addition, $|z|\to\iy$ and $z\in \cZ_+^+$.
Then
\[
\lb{asdphiS}
\xi (z)
=\det\ma
\phi_1(0,z)&\phi_2(0,z)\\
\phi_1(1,z)&\phi_2(1,z)
\am+O(e^{-\Re z}).
\]

ii) The eigenvalues $\m_n$ satisfy
\[
\lb{asmnrh}
\m_n=\big(\n n+O(n^{-1})\big)^3,
\]
as $n\to+\iy,\n={2\pi\/\sqrt3}$.

\end{lemma}

\no {\bf Proof.}
i) Let $|z|\to\iy,z\in \cZ$. The asymptotics \er{asphij} and the definitions \er{4g.Om} give
$$
|\phi_1(x,z)|=e^{(-\Re z-\sqrt 3\Im z){x\/2}}\big(1+O(z^{-1})\big),\qq
|\phi_2(x,z)|=e^{(-\Re z+\sqrt 3\Im z){x\/2}}\big(1+O(z^{-1})\big).
$$
Let $|z|\to\iy,z\in \cZ_+^+$.
Then $\Im z\le {\Re z\/\sqrt3}$ and substituting these asymptotics
into the definition \er{defphiz} we obtain
$$
\det\phi(z)=\ma
\phi_1(0,z)&\phi_2(0,z)&O(1)\\
\phi_1(1,z)&\phi_2(1,z)&e^{\Re z}O(1)\\
e^{-\Re z-\sqrt3\Im z}O(1)&e^{-\Re z+\sqrt3\Im z}O(1)
&\phi_3(2,z)
\am,
$$
The asymptotics \er{asmphi3} implies
$$
\xi (z)=\ma
\phi_1(0,z)&\phi_2(0,z)&e^{-2\Re z}O(1)\\
\phi_1(1,z)&\phi_2(1,z)&e^{-\Re z}O(1)\\
e^{-\Re z-\sqrt3\Im z}O(1)&e^{-\Re z+\sqrt3\Im z}O(1)
&1
\am,
$$
which yields \er{asdphiS}.

ii)  Let $\l=z^3=\m_n,n\to+\iy$.
Lemma~\ref{CLev}~ii) yields $z=\n n+\d_n,\d_n=O(1)$
and $\d_n$ is real.
The asymptotics \er{asphij} implies
$$
\phi_1(0,z)=1+O(n^{-1}),\qq \phi_2(0,z)=1+O(n^{-1}),
$$
$$
\phi_1(1,z)
=e^{{-1+i\sqrt3\/2} (\n n+\d_n)}\big(1+O(n^{-1})\big)
=(-1)^ne^{-{\n n\/2}}\Big(1-{1-i\sqrt3\/2}\d_n+O(\d_n^2)+O(n^{-1})\Big),
$$
$$
\phi_2(1,z)
=e^{{-1-i\sqrt3\/2} (\n n+\d_n)}\big(1+O(n^{-1})\big)
=(-1)^ne^{-{\n n\/2}}\Big(1-{1+i\sqrt3\/2}\d_n+O(\d_n^2)+O(n^{-1})\Big).
$$
Substituting these asymptotics into \er{asdphiS} we obtain
$$
\xi (z)=(-1)^ne^{-{\n n\/2}}\big(-i\sqrt3\d_n+O(\d_n^2)+O(n^{-1})\big).
$$
The identity $\xi (z)=0$, see \er{detAev}, gives $\d_n=O(n^{-1})$,
which yields \er{asmnrh}.
\BBox

\subsection{Eigenvalue asymptotics}
Now we determine eigenvalue asymptotics for the case $p,q\in L^1(\T)$.
Introduce the Fourier coefficients
$$
f_0=\int_0^1f(x)dx,\qqq \wh f_n=\int_0^1 e^{-i2\pi nx}f(x)dx,\qqq n\in\Z.
$$
Note that the coefficients, given by \er{wtFc}, satisfy
$$
\wt f_n=\Re\wh f_n-{\Im\wh f_n\/\sqrt3},\qq n\in\N.
$$

\begin{lemma}
\lb{Lmevns}
Let $p,q\in L^1(\T)$. Then the eigenvalues
$\m_n$ satisfy the asymptotics \er{asmn1}.
\end{lemma}

\no {\bf Proof.}
Let $p,q\in L^1(\T)$ and let the matrix-valued function $ A$ be given by
\er{idPhi0}. Then $\cW=0$ in the asymptotics \er{idPhiGen} and
in the considered case we have
$$
m=1,\qq \cF=-{p\/3}(P-2\cT^2),\qq \Theta=\cT-{2p\/3z^2}\cT^2,
$$
see \er{vnsc} and \er{4g.wtOss}.
The identity \er{phi1jpr} gives
\[
\lb{phi1j}
\zeta_{j}(x,z)={\eta_j(x,z)\/z},\qqq j=1,2,3,
\]
where $\eta_j$ are given by \er{defcGj}.
Let
$$
\l=z^3=\m_n,\qqq n\to+\iy.
$$
Lemma~\ref{CLev}~v)  shows $z={2\pi n\/\sqrt 3}+\d_n,\d_n=O(n^{-1})$.
Substituting the asymptotics \er{idG1f} into the identity \er{phi1j}
and using the definitions \er{cK12} we obtain
$$
\begin{aligned}
\zeta_{1}(x,z)={1\/3z}\int_x^2e^{i2\pi n(s-x)}p(s)P_{21}ds+O(n^{-{3\/2}})
=-{\t^2\/3z}\int_x^2e^{i2\pi n(s-x)}p(s)ds+O(n^{-{3\/2}}),\\
\zeta_{2}(x,z)=-{1\/3z}\int_0^xe^{i2\pi n(x-s)}p(s)P_{12}ds+O(n^{-{3\/2}})
={\t\/3z}\int_0^xe^{i2\pi n(x-s)}p(s)ds+O(n^{-{3\/2}}).
\end{aligned}
$$
Then
$$
\begin{aligned}
\zeta_{1}(0,z)=-{2\t^2\/3z}\ol{\wh p_n}+O(n^{-{3\/2}}),\qq
\zeta_{2}(0,z)=O(n^{-{3\/2}}),\\
\zeta_{1}(1,z)=-{\t^2\/3z}\ol{\wh p_n}+O(n^{-{3\/2}}),\qq
\zeta_{2}(1,z)={\t\/3z}\wh p_n+O(n^{-{3\/2}}),
\end{aligned}
$$
and the asymptotics \er{asphijepr} gives
$$
\phi_1(0,z)=1-{2\t^2\/3z}\ol{\wh p_n}+O(n^{-{3\/2}}),
\qq
\phi_2(0,z)=1+O(n^{-{3\/2}}),
$$
$$
\phi_1(1,z)=e^{\t z-{2p_0\/3z^2}\t^2}
\Big(1-{\t^2\/3z}\ol{\wh p_n}+O(n^{-{3\/2}})\Big),
\qq
\phi_2(1,z)=e^{\t^2z-{2p_0\/3z^2}\t}\Big(1+{\t\/3z}\wh p_n
+O(n^{-{3\/2}})\Big).
$$
Substituting these asymptotics into \er{asdphiS}
and using the identities
$$
e^{\t z}=e^{-{z\/2}+i{\sqrt3\/2} z}=(-1)^ne^{-{z\/2}}e^{i{\sqrt3\/2} \d_n},\qq
e^{\t^2 z}=(-1)^ne^{-{z\/2}}e^{-i{\sqrt3\/2} \d_n}
$$
 we obtain
\[
\lb{Axi1}
\xi (z)=(-1)^ne^{-{z\/2}+{p_0\/3z}}\big(\x_n+O(e^{-{\pi n\/\sqrt3}})\big),
\]
where
$$
\x_n=\det\ma 1
-{2\t^2\/3z}\ol{\wh p_n}+O(n^{-{3\/2}})&1+O(n^{-{3\/2}})
\\
e^{i{\sqrt3\/2} \d_n+i{p_0\/\sqrt3 z}}\big(1-{\t^2\/3z}\ol{\wh p_n}+O(n^{-{3\/2}})\big)&
e^{-i{\sqrt3\/2} \d_n-i{p_0\/\sqrt3 z}}\big(1+{\t\/3z}\wh p_n+O(n^{-{3\/2}})\big)\am.
$$
Using the asymptotics $e^{\pm i{\sqrt3\/2} \d_n}=1\pm i{\sqrt3\/2} \d_n+O(n^{-2})$ we obtain
$$
\begin{aligned}
\x_n=\det\ma 1
-{2\t^2\/3z}\ol{\wh p_n}+O(n^{-{3\/2}})&1+O(n^{-{3\/2}})
\\
1+i{\sqrt3\/2} \d_n+i{p_0\/\sqrt3 z}
-{\t^2\/3z}\ol{\wh p_n}+O(n^{-{3\/2}})&
1-i{\sqrt3\/2} \d_n-i{p_0\/\sqrt3 z}+{\t\/3z}\wh p_n
+O(n^{-{3\/2}})\am
\\
=-i\sqrt3\d_n-{i2 p_0\/\sqrt3z}
+{\t\wh p_n-\t^2\ol{\wh p_n}\/3z}
+O(n^{-{3\/2}})
=-i\sqrt3\Big(\d_n+{2 p_0\/3z}
-{2\Im(\t\wh p_n)\/3\sqrt3z}\Big)
+O(n^{-{3\/2}}).
\end{aligned}
$$
The identities \er{detAev} and \er{Axi1} imply $\x_n=O(e^{-{\pi n\/\sqrt3}})$, which gives
$$
\d_n=-{2 p_0\/3z}
+{2\Im(\t\wh p_n)\/3\sqrt3z}+O(n^{-{3\/2}})
=-{p_0\/\sqrt3\pi n}
+{\Im(\t\wh p_n)\/3\pi n}+O(n^{-{3\/2}}).
$$
Using the identity $\Im(\t\wh p_n)={1\/2}(\sqrt3\Re\wh p_n-\Im\wh p_n)$
we obtain
$$
z=\n n-{p_0\/\sqrt3\pi n}
+{\sqrt3\Re\wh p_n-\Im\wh p_n\/6\pi n}+O(n^{-{3\/2}}).
$$
which yields the asymptotics \er{asmn1} as $n\to+\iy$.

This identity \er{symev} and $\wh{p_n^-}=\wh p_{-n}$
give the asymptotics \er{asmn1} for $n\to-\infty$.
\BBox

\section{The case $p',q\in L^1(\T)$}
\setcounter{equation}{0}
\lb{Sect6}

\subsection{Transformation of the differential equation}
Integrating by part in the integral operator $K$
of the integral equation \er{4g.me5ipr}, given by the definition \er{4g.dcLipr},
and repeating the previous analysis we obtain
the asymptotics for the case of smooth coefficients.
However, in order to simplify the calculations
we use the other method
based on an additional transformation
of the differential equation, see  Fedoryuk \cite[Ch~V.1.3]{F12}.

Let $p',q\in L^1(\T)$, let $r>0$ be large enough and let $z\in \cZ_+(r)$.
Introduce the
matrix-valued function $U_1(x,z),x\in\R$ by
\[
\lb{defcU1}
U_1(x,z)=\1_3+{p(x)\/3z^2}W_1,
\]
where
\[
\lb{defW1ss}
W_1={i\/\sqrt3}\ma 0&\t&-\t\\-\t^2&0&\t^2\\1&-1&0\am,
\]
The matrix $U_1(x,z)$ is invertible at large $|z|$ and we
introduce the matrix $Y_1(x,z)$ by the identity
\[
\lb{cMFWss}
Y(x,z)=U_1(x,z)Y_1(x,z)U_1^{-1}(0,z),
\qqq (x,z)\in\R\ts \cZ_+(r),
\]
where $Y$ the solution of the problem \er{eqcM}.

\begin{lemma}
\lb{Lmfm2}
Let $p',q\in L^1(\T)$ and let $z\in \cZ_+(r)$,
where $r>0$ is large enough.  Then
the matrix-valued function $ Y_1(x,z)$,
given by \er{cMFWss}, satisfies the equation
\[
\lb{me2ss}
 Y_1'-zT_1 Y_1
={1\/z^2}\Phi_1 Y_1,\qqq Y_1(0,z)=\1_3,
\]
where
$$
\Phi_1(x,z)=P_1(x)-{p^2(x)\/9z}(PW_1-2W_1\cT^2)+O(z^{-2})
$$
the asymptotics is uniform on $x\in\R$,
$T_1$ is given by the definition \er{4g.wtOss} and
\[
\lb{defP1}
P_1(x)=-{1\/3}\big(q(x)Q+p'(x)W_1\big).
\]

\end{lemma}

\no {\bf Proof.}
Let $z\in \cZ_+(r)$.
Substituting \er{cMFWss} into equation \er{eqcM} we obtain
\[
\lb{me11ss}
 Y_1'=U_1^{-1}A_1 Y_1,\qq
A_1(x,z)=\Big(z\cT -{p(x)\/3z} P-{q(x)\/3z^2} Q\Big)U_1(x,z)-{p'(x)\/3z^2}W_1.
\]
Let $|z|\to\iy$.
Then
$$
A_1(x,z)=z\cT +{p(x)\/3z}(\cT W_1-P)+{P_1(x)\/z^2}-{p^2(x)\/9z^3}PW_1 +O(z^{-4}),
$$
and
$$
 U_1^{-1}(x,z)=\1_3-{p(x)\/3z^2}W_1+{p^2(x)\/9z^4}W_1^2+O(z^{-6}).
$$
uniformly in $x$.
We have
\[
\lb{W-1Bbss}
\begin{aligned}
 U_1^{-1}(x,z)A_1(x,z)
=z\cT -{p(x)\/3z}\big([ W_1,\cT] +P\big)
+{P_1(x)\/z^2}
\\
+{p^2(x)\/9z^3}\Big(W_1[W_1,\cT]+[W_1,P]\Big)
+O(z^{-4}),
\end{aligned}
\]
here and below $[A,B]=AB-BA$.
The matrix $W_1$ satisfies
\[
\lb{cOW1ss}
[W_1,\cT]+P=\diag(P_{jj})_{j=1}^3=2\cT^2.
\]
Substituting \er{cOW1ss}  into \er{W-1Bbss}
we obtain
$$
 U_1^{-1}(x,z)A_1(x,z)
=z\cT -{2p(x)\/3z}\cT^2+{P_1(x)\/z^2}
-{p^2(x)\/9z^3}(PW_1-2W_1\cT^2)
+O(z^{-4}).
$$
Substituting this identity into equation \er{me11ss}
we obtain \er{me2ss}.
\BBox

\subsection{Factorization of the fundamental matrix}
Now we apply Theorem~\ref{Thf} to equation \er{me2ss}
in order to obtain the factorization of the fundamental matrix.
Preliminary, as well as in the case of equation \er{eqcM} (see Sect~\ref{Sectffm0})
we have to replace the diagonal part $-{q\/3}\cT$ of the matrix
$P_1$ in the right side of \er{me2ss} into the left one in order to get
the off-diagonal matrix $\cF$ in the asymptotics of $\Phi$ in \er{ascQTh}.
Then we obtain the following corollary of Theorem~\ref{Thf}
on the factorization of the fundamental matrix.

\begin{corollary}

Let $p',q\in L^1(\T)$ and let $z\in \cZ_+(r)$,
where $r>0$ is large enough.  Then
the fundamental matrix $M$ has the representation
$M(x,\l)= A(x,z) A^{-1}(0,z)$, where $\l=z^3$
and the matrix-valued function $ A$ has the form
\[
\lb{idPhiss}
 A(x,z)=\Omega(z) U_1(x,z)\cX(x,z)
e^{z\int_0^xT_2(s,z)ds},\qq (x,z)\in [0,2]\ts \cZ_+(r).
\]
\[
\lb{4g.wtO}
T_2=\cT-{2p\/3z^2}\cT^2-{q\/3z^3}\cT,
\]
and $\cX$ is the solution of the integral equation \er{4g.me5ipr}
with
\[
\lb{vnscss}
m=2,\qq\Theta=T_2,\qq \Phi=\Phi_1+{q\/3}\cT.
\]
Each function $ A(x,\cdot),x\in[0,2]$, is analytic in $\cZ_+(r)$, where
$r>0$ is large enough.

\end{corollary}

\no {\bf Proof.}
Rewrite equation \er{me2ss} in the form
\[
\lb{eqcM1}
 Y_1'-zT_2 Y_1
={1\/z^2}\Big(\Phi_1+{q\/3}\cT\Big) Y_1,
\]
Equation \er{eqcM1} has the form \er{me2pr} with $m,\Theta,\Phi$
satisfying \er{vnscss}.
Let $ Y_1$ be the solution of equation \er{eqcM1} satisfying the condition
$ Y_1(0)=\1_3$.
Theorem~\ref{Thf} shows that $ Y_1$ satisfies the identity \er{4g.rmmipr}.
The identities \er{defcM} and \er{cMFWss}
imply the identity \er{MsimPhi}, where $ A$ satisfies \er{idPhiss}.
\BBox

\subsection{Eigenvalue asymptotics}
Introduce the function $h(x),x\in\T$, and the constant $\vk$ by
\[
\lb{defh}
h=\t(p'-i\sqrt3 q),\qqq \vk=i(\t-1).
\]

\begin{lemma}
Let $p',q\in L^1(\T)$ and let $n\to\pm\iy$. Then the eigenvalues $\m_n$ satisfy
\[
\lb{asmnss}
\m_n=\m_{n}^o-2\n np_0+q_0 +\Re\Big({\wh{p'_n}\/3}-\wh
q_n\Big)+{\Im(\wh q_n+\wh{p'_n})\/\sqrt3} +O(n^{-{1\/2}}).
\]
\end{lemma}

\no {\bf Proof.}
Let $p',q\in L^1(\T)$. Without loss of generality we assume that
$\int_0^1 q(t)dt=0$. Let the matrix-valued function $ A$ be given by
\er{idPhiss}. Then $\cW={p\/3}W_{1}$ in the asymptotics \er{idPhiGen} and,
due to \er{vnscss}, in the considered case we have
$$
m=2,\qq\cF=P_1+{q\/3}\cT,\qq\Theta=\cT-{2p\/3z^2}\cT^2-{q\/3z^3}\cT,
$$
$P_1$ is given by \er{defP1}.
The identity \er{phi1jpr} gives
\[
\lb{phi1jss}
\zeta_{j}(x,z)={\eta_{j}(x,z)\/z^2}+{p(x)\eta_{1,j}\/3z^2},\qqq
\eta_{1,j}=\sum_{m =1}^3W_{1,mj},
\]
and $\eta_j$ are given by the definition \er{defcGj}.
The definition \er{defW1ss} gives
\[
\lb{idW1}
\eta_{1,1}={\ol\vk\/\sqrt3},\qqq\eta_{1,2}={\vk\/\sqrt3}.
\]
Let $\l=z^3=\m_n,n\to+\iy$.
The asymptotics \er{asmn1} show that
$
z=\n n-{p_0\/\sqrt3\pi n}+\d_n,\d_n=o(n^{-1}).
$
The identity \er{idG1f} and the definitions  \er{cK12} and  \er{defW1ss} imply
$$
\begin{aligned}
\eta_1(x,z)=-\int_x^2e^{i2\pi n(s-x)}P_{1,21}(s)ds+O(n^{-{1\/2}})
=-{i\/3\sqrt3} \int_x^2e^{i2\pi n(s-x)}\ol h(s)ds+O(n^{-{1\/2}}),
\\
\eta_2(x,z)=\int_0^xe^{i2\pi n(x-s)}P_{1,12}(s)ds+O(n^{-{1\/2}})
=-{i\/3\sqrt3} \int_0^xe^{i2\pi n(x-s)}h(s)ds+O(n^{-{1\/2}}).
\end{aligned}
$$
where we used the identities
$$
P_{1,21}={i\ol h\/3\sqrt3},\qqq P_{1,12}=-{ih\/3\sqrt3}.
$$
Substituting these asymptotics into the identity \er{phi1jss}
and using the identities \er{idW1}
we obtain
$$
\zeta_{1}(x,z)={\ol\vk p(x)\/3\sqrt3 z^2}
-{i \/3\sqrt3z^2}\int_x^2e^{i2\pi n(s-x)}\ol h(s)ds
+O(n^{-{5\/2}}),
$$
$$
\zeta_{2}(x,z)={\vk p(x)\/3\sqrt3z^2}
-{i\/3\sqrt3z^2}\int_0^xe^{i2\pi n(x-s)}h(s)ds
+O(n^{-{5\/2}}).
$$
Then the asymptotics \er{asphijepr} gives
$$
\phi_1(0,z)=1+{\ol\vk p(0)-2i\ol{\wh h_n}\/ 3\sqrt3z^2}+O(n^{-{5\/2}}),
\qq
\phi_1(1,z)=e^{\t z-{2p_0\/3z}\t^2}
\Big(1+{\ol\vk p(0)-i\ol{\wh h_n}\/ 3\sqrt3z^2}
+O(n^{-{5\/2}})\Big),
$$
$$
\phi_2(0,z)=1+{\vk p(0)\/3\sqrt3z^2}+O(n^{-{5\/2}}),
\qq
\phi_2(1,z)=e^{\t^2 z-{2p_0\/3z}\t}
\Big(1+{\vk p(0)-i\wh h_n\/3\sqrt3z^2}
+O(n^{-{5\/2}})\Big)
$$
where
we used \er{4g.wtOss}.
Substituting these asymptotics into \er{asdphiS}
and using the identities
\[
\lb{ideomzim}
e^{\t z-{2p_0\/3z}\t^2}
=(-1)^ne^{-{z\/2}+{p_0\/3z}}e^{i{\sqrt3\/2} \d_n},\qq
e^{\t^2 z-{2p_0\/3z}\t}=(-1)^ne^{-{z\/2}+{p_0\/3z}}e^{-i{\sqrt3\/2} \d_n}
\]
 and $\int_0^1q(s)ds=0$ we obtain
$$
\xi (z)=(-1)^ne^{-{z\/2}+{p_0\/3z}}\big(\x_n+e^{-{\pi n\/\sqrt3}}O(1)\big),
$$
where
$$
\begin{aligned}
\x_n
=\det\ma1+{\ol\vk p(0)-2i\ol{\wh h_n}\/ 3\sqrt3z^2}
+O(n^{-{5\/2}})&
1+{\vk p(0)\/3\sqrt3z^2}+O(n^{-{5\/2}})\\
1+i{\sqrt3\/2} \d_n
+{\ol\vk p(0)-i\ol{\wh h_n}\/ 3\sqrt3z^2}
+O(n^{-{5\/2}})&
1-i{\sqrt3\/2} \d_n
+{\vk p(0)-i\wh h_n\/3\sqrt3z^2}
+O(n^{-{5\/2}})\am
\\
=-i\sqrt3 \d_n
-{i(\wh h_n+\ol{\wh h_n})\/3\sqrt3z^2}
+O(n^{-{5\/2}})=-i\sqrt3\d_n
-{i2\Re\wh h_n\/\sqrt3(2\pi n)^2}
+O(n^{-{5\/2}}).
\end{aligned}
$$
The identity \er{detAev} implies $\x_n+e^{-{\pi n\/\sqrt3}}O(1)=0$, which gives
\[
\lb{asdnpr}
\d_n=-{2\Re\wh h_n\/3(2\pi n)^2}+O(n^{-{5\/2}}).
\]
The definition \er{defh} gives
$$
\Re\wh h_n=\sqrt3\Im(\t\wh q_n)+\Re(\t\wh{p'_n}).
$$
Using the identities
$$
\Im(\t\wh q_n)={1\/2}(\sqrt3\Re\wh q_n-\Im\wh q_n),\qq
\Re(\t\wh{p'_n})=-{1\/2}(\sqrt3\Im\wh{p'_n}+\Re\wh{p'_n})
$$
we obtain
$$
\Re\wh h_n={\sqrt3\/2}\Big(\sqrt3\Re\big(\wh q_n-{\wh{p'_n}\/3}\big)
-\Im(\wh{p'_n}+\wh q_n)\Big).
$$
Substituting this identity into \er{asdnpr} we obtain
$$
\d_n=-{1\/(2\pi n)^2}\Big(\Re\big(\wh q_n-{\wh{p'_n}\/3}\big)
-{\Im(\wh{p'_n}+\wh q_n)\/\sqrt3}\Big)+O(n^{-{5\/2}}),
$$
which gives
$$
z=\n n-{p_0\/\sqrt3\pi n}
-{1\/(2\pi n)^2}\Big(\Re\big(\wh q_n-{\wh{p'_n}\/3}\big)
-{\Im(\wh{p'_n}+\wh q_n)\/\sqrt3}\Big)+O(n^{-{5\/2}}).
$$
This yields the asymptotics \er{asmnss} as $n\to+\iy$.
The identities \er{symev} and
$\wh{q_n^-}=\wh q_{-n},\wh{(p^-)'_n}=-\wh p'_{-n}$
give the asymptotics \er{asmnss} for $n\to-\infty$.
\BBox

\section{The case $p'',q'\in L^1(\T)$}
\setcounter{equation}{0}
\lb{Sect7}

\subsection{Transformation of the differential equation}
In the previous Section we determined eigenvalue asymptotics for the case
$p',q\in L^1(\T)$.  Using the similar arguments in this Section
we determine eigenvalue asymptotics for the case
$p'',q'\in L^1(\T)$.
First of all, in order to avoid the difficulties arising from
the application of integration by parts, we transform the differential equation.

Let $p'',q'\in L^1(\T)$. Let $r>0$ be large enough and let $z\in \cZ_+(r)$.
Introduce the matrix-valued function
\[
\lb{defcU2}
 U_2(x,z)= U_1(x,z)+{W_2(x)\/3z^3}=\1_3+{p(x)W_1\/3z^2}+{W_2(x)\/3z^3},\qq
x\in\R,
\]
where
\[
\lb{defW2}
W_2={1\/3}\ma 0& h&\ol h\\
\ol h&0&h\\
h&\ol h&0\am,
\]
$h$ is given by \er{defh}.
The matrix $ U_2(x,z)$ is invertible at large $|z|$ and we
introduce the matrix $ Y_2(x,z)$ by the identity
\[
\lb{cMFW}
Y(x,z)= U_2(x,z) Y_2(x,z)
 U_2^{-1}(0,z),
\qq (x,z)\in\R\ts \cZ_+(r),
\]
where $Y$ is the solution of the problem \er{eqcM}.

\begin{lemma}
\lb{Lmfm3}
Let $p'',q'\in L^1(\T)$ and let $z\in \cZ_+(r)$,
where $r>0$ is large enough.  Then
the matrix-valued function $ Y_2(x,z)$,
given by \er{cMFW}, satisfies the equation
\[
\lb{me2}
 Y_2'-zT_2 Y_2={1\/z^3}\Phi_2 Y_2,\qqq
 Y_2(0,z)=\1_3,
\]
where $T_2$ is given by the definition \er{4g.wtO},
\[
\lb{decQ}
\Phi_2(x,z)=-{1\/3}W_2'(x)-{p^2(x)\/9}(PW_1-2W_1\cT^2)+O(z^{-1}),
\]
uniformly on $x\in\R$.

\end{lemma}

\no {\bf Proof.}
Let $(x,z)\in\R\ts \cZ_+(r)$.
The definitions \er{cMFWss} and \er{cMFW} give
$$
 Y_1(x,z)= U_3(x,z) Y_2(x,z) U_3^{-1}(0,z),
\qqq  U_3= U_1^{-1} U_2.
$$
Substituting this identity into equation \er{me2ss} we obtain
\[
\lb{me2ss1}
 Y_2' = U_3^{-1}\lt(\Big(zT_1
+{P_1\/z^2}-{p^2\/9z^3}Q_1\Big) U_3
- U_3'+O(z^{-4})\rt) Y_2,
\]
where
$$
P_1=-{1\/3}(qQ+p'W_1),\qq Q_1=PW_1-2W_1\cT^2.
$$
The definitions \er{defcU1} and \er{defcU2} imply
$$
 U_3(x,z)=\1_3+{W_2(x)\/3z^3}+O(z^{-5}),\qqq
 U_3'(x,z)={W_2'(x)\/3z^3}+O(z^{-5}).
$$
Substituting this asymptotics into equation \er{me2ss1} we obtain
\[
\lb{me11}
 Y_2'
=\Big(zT_1+{1\/3z^2}\big(3P_1+[\cT, W_2]\big)
-{1\/9z^3}\big(p^2Q_1+3W_2'\big)+O(z^{-4})\Big) Y_2,
\]
where we used the definition \er{4g.wtOss}.
The matrix $W_2$ satisfies
$$
[\cT, W_2]+3P_1=-q\cT.
$$
Substituting this identity into equation \er{me11}
we obtain
$$
 Y_2'
=\Big(zT_1-{q\cT\/3z^2}
-{1\/9z^3}\big(p^2Q_1+3W_2'\big)+O(z^{-4})\Big) Y_2.
$$
The identity $T_2=T_1-{q\/3z^3}\cT$ gives \er{me2}.
\BBox

\subsection{Factorization of the fundamental matrix}
We apply Theorem~\ref{Thf} to equation \er{me2}
in order to obtain the factorization of the fundamental matrix.
In this case the matrix
$-{1\/3}W_2'(x)-{1\/9}p^2(x)(PW_1-2W_1\cT^2)$
in the right side is off-diagonal and we can apply
Theorem~\ref{Thf} immediately to equation \er{me2}.
Then we obtain the following corollary of Theorem~\ref{Thf}
on the factorization of the fundamental matrix.

\begin{corollary}
Let $p'',q'\in L^1(\T)$ and let $z\in \cZ_+(r)$,
where $r>0$ is large enough.  Then the fundamental matrix $M$ has the representation
$M(x,\l)= A(x,z) A^{-1}(0,z)$, where  $\l=z^3$
and the matrix-valued function $ A$ has the form
\[
\lb{idPhi}
 A(x,z)=\Omega(z)  U_2(x,z)\cX(x,z)
e^{z\int_0^xT_2(s,z)ds},\qq (x,z)\in [0,2]\ts \cZ_+(r),
\]
and $\cX$ is the solution of the integral equation \er{4g.me5ipr}
with
\[
\lb{mwth}
m=3,\qq\Theta=T_2,\qq \Phi=\Phi_2.
\]
Each function $ A(x,\cdot),x\in[0,2]$, is analytic in $\cZ_+(r)$, where
$r>0$ is large enough.
\end{corollary}

\no {\bf Proof.}
Equation \er{me2} has the form \er{me2pr} with $m,\Theta,\Phi$
satisfying \er{mwth}.
Let $ Y_2$ be the solution of equation \er{me2} satisfying the condition
$ Y_2(0)=\1_3$.
Theorem~\ref{Thf} shows that $ Y_2$ satisfies the identity \er{4g.rmmipr}.
The identities \er{defcM}, \er{cMFW}
imply the identity \er{MsimPhi}, where $ A$ satisfies \er{idPhi}.
\BBox

\subsection{Eigenvalue asymptotics}
Now we determine the eigenvalue asymptotics for the case
$p'',q'\in L^1(\T)$.

\begin{lemma}
Let $p'',q'\in L^1(\T)$ and let $n\to\pm\iy$. Then
the eigenvalues $\m_n$ satisfy
\[
\lb{asmn}
\m_n=\m_{n}^o-2\n np_0+q_0
+{2p_0^2\/\sqrt3\pi n}
+{1\/2\pi n}\Big(\Im\Big({\wh{p_n''}\/3}-\wh{q_n'}\Big)
-{\Re(\wh{p_n''}+\wh{q_n'})\/\sqrt3}\Big)
+O(n^{-{3\/2}}).
\]
\end{lemma}

\no {\bf Proof.}
Let $p'',q'\in L^1(\T)$. Without loss of generality we assume that
$\int_0^1 q(t)dt=0$. Let the matrix-valued function $ A$ be given by
\er{idPhi}. Then $\cW(x)={p\/3}W_{1}+{1\/3z}W_2$
in the asymptotics \er{idPhiGen} and,
due to  Lemma~\ref{Lmfm3}~ii), in the considered case we have
$$
m=3,\qq \cF=-{1\/3}\Big(W_2'+{p^2\/3}Q_1\Big),\qq
\Theta=\cT-{2p\/3z^2}\cT^2-{q\/3z^3}\cT,
$$
where $Q_1=PW_1-2W_1\cT^2$.
The identity \er{phi1jpr} gives
\[
\lb{phi1jsc}
\zeta_{j}(x,z)={p(x)\eta_{1,j}\/3z^2}+{ \eta_{2,j}(x)\/3z^3}
+{\eta_{j}(x,z)\/z^3},\qq \eta_{2,j}(x)=\sum_{m =1}^3W_{2,mj}(x),
\]
where $\eta_{1,j}$ are given by \er{phi1jss}.
The definition \er{defW2} implies
\[
\lb{idW2}
\eta_{2,1}(x)=\eta_{2,2}(x)={1\/3}(h+\ol h)=q-{p'\/3}=V,
\]
where $h$ is given by \er{defh} and we used \er{deffh}.
Let $\l=z^3=\m_n,n\to+\iy$. Integrating by parts in the asymptotics \er{asmnss} we obtain
$$
z=\n n-{p_0\/\sqrt3\pi n}+\d_n,\qqq \d_n=O(n^{-{5\/2}}).
$$
The definition
\er{defW2} provides
\[
\lb{idmR1}
\cF_{12}=-{h' \/9}-{p^2\/9}Q_{1,12},\qq
\cF_{21}=-{\ol{h'}\/9}-{p^2\/9}Q_{1,21}
\]
Substituting this identities into \er{idG1f} and using the asymptotics
$$
\int_0^xe^{-i2\pi ns}p^2(s)=O(n^{-1}),\qqq
\int_x^2e^{-i2\pi ns}p^2(s)=O(n^{-1}),
$$
we obtain
$$
\begin{aligned}
\eta_1(x,z)={1\/9}\int_x^2e^{i2\pi n(s-x)}\ol{h'}(s)ds+O(n^{-{1\/2}}),
\\
\eta_2(x,z)=-{1\/9}\int_0^xe^{i2\pi n(x-s)}h'(s)ds+O(n^{-{1\/2}}).
\end{aligned}
$$
Substituting this asymptotics into the definition \er{phi1jsc}
and using the identities \er{idW1} and \er{idW2}
we obtain
$$
\zeta_{1}(x,z)={\ol\vk p(x)\/3\sqrt3z^2}
+{V(x)\/3z^3}
+{1\/9z^3}\int_x^2e^{i2\pi n(s-x)}
\ol{h'}(s)ds+O(n^{-{7\/2}}),
$$
$$
\zeta_{2}(x,z)={\vk p(x)\/3\sqrt3z^2}+{V(x)\/3z^3}
-{1 \/9z^3}\int_0^xe^{i2\pi n(x-s)}
h'(s)ds+O(n^{-{7\/2}}),
$$
where $\vk$ is given by \er{defh}.
Then the asymptotics \er{asphijepr} gives
\[
\lb{asphi101}
\phi_1(0,z)=1+\b(z)+{2\ol{\wh{h_n'}}\/9z^3}
+O(n^{-{7\/2}}),
\qq
\phi_1(1,z)=e^{z(\t-{2p_0\t^2\/3z^2})}
\Big(1+\b(z)+{\ol{\wh{h_n'}}\/9z^3}
+O(n^{-{7\/2}})\Big),
\]
\[
\lb{asphi201}
\phi_2(0,z)=1+\a(z)
+O(n^{-{7\/2}}),
\qq
\phi_2(1,z)=e^{z(\t^2-{2p_0\t\/3z^2})}
\Big(1+\a(z)-{\wh{h_n'}\/9z^3}
+O(n^{-{7\/2}})\Big),
\]
where
$$
\a(z)={\vk p(0)\/3\sqrt3 z^2}+{V(0)\/3z^3},\qqq
\b(z)={\ol\vk p(0)\/3\sqrt3 z^2}+{V(0)\/3z^3}
$$
and we used \er{4g.wtO} and the identity $\int_0^1q(s)ds=0$.
Substituting the asymptotics \er{asphi101} and \er{asphi201}
into \er{asdphiS} and using \er{ideomzim} and the asymptotics $\d_n=O(n^{-{5\/2}})$
 we obtain
$$
\xi (z)=(-1)^ne^{-{z\/2}+{p_0\/3z}}
\Big(\x_n+e^{-{\pi n\/\sqrt3}}O(1)\Big),
$$
where
\[
\lb{asBzevsc}
\begin{aligned}
\x_n=\det\ma
1+\b(z)+{2\ol{\wh{h_n'}}\/9z^3}
+O(n^{-{7\/2}})&1+\a(z)
+O(n^{-{7\/2}})\\
1+i{\sqrt3\/2} \d_n
+\b(z)+{\ol{\wh{h_n'}}\/9z^3}
+O(n^{-{7\/2}})&1-i{\sqrt3\/2} \d_n
+\a(z)-{\wh{h_n'}\/9z^3}
+O(n^{-{7\/2}})
\am
\\
=-i\sqrt3 \d_n
+{\ol{\wh{h_n'}}-\wh{h_n'}\/9z^3}+O(n^{-{7\/2}})=-i\sqrt3 \d_n
-{2i\Im\wh{h_n'}\/\sqrt3(2\pi n)^3}+O(n^{-{7\/2}}).
\end{aligned}
\]
The identity \er{detAev} implies $\x_n+e^{-{\n n\/2}}O(1)=0$, which gives
\[
\lb{asdnprsc}
\d_n=-{2\Im\wh{h_n'}\/3(2\pi n)^3}+O(n^{-{7\/2}}).
\]
The definition \er{defh} gives
$$
\Im\wh{h_n'}=\Im\big(\t \wh{p_n''}\big)-\sqrt3\Re\big(\t \wh{q_n'}\big).
$$
Substituting this identity into \er{asdnprsc} and using
$$
\Im\big(\t\wh{p_n''}\big)={1\/2}\big(\sqrt3\Re\wh{p_n''}-\Im\wh{p_n''}\big),\qq
\Re(\t\wh{q_n'})={1\/2}\big(-\sqrt3\Im\wh{q_n'}-\Re\wh{q_n'}\big),
$$
we obtain
$$
\d_n={1\/3(2\pi n)^3}\Big(
\sqrt3\big(-\sqrt3\Im\wh{q_n'}-\Re\wh{q_n'}\big)
-\big(\sqrt3\Re\wh{p_n''}-\Im\wh{p_n''}\big)
\Big)+O(n^{-{7\/2}}),
$$
which gives
$$
z=\n n-{p_0\/\sqrt3\pi n}+
{1\/3(2\pi n)^3}\Big(
\sqrt3\big(-\sqrt3\Im\wh{q_n'}-\Re\wh{q_n'}\big)
-\big(\sqrt3\Re\wh{p_n''}-\Im\wh{p_n''}\big)
\Big)+O(n^{-{7\/2}}).
$$
This yields the asymptotics \er{asmn} as $n\to+\iy$.
The identities \er{symev} and
$\wh{(q^-)'_n}=-\wh q'_{-n},\wh{(p^-)''_n}=\wh p''_{-n}$
give the asymptotics \er{asmn} for $n\to-\infty$.
\BBox

\medskip
\no {\bf Proof of Theorem~\ref{Thevas}.}
Lemma~\ref{CLev}~iii) shows that
there exists $N\in\N$ such that each $\m_n,|n|>N$, is real.
The asymptotics \er{asmn1} is proved in Lemma~\ref{Lmevns}.
Substituting the identities
$$
\wh{p'_n}=\int_0^1e^{-i2\pi nx}p'(x)dx=i2\pi n\wh{p_n},\qq
\wh{q'_n}=i2\pi n\wh{q_n},\qq
\wh{p''_n}=-(2\pi n)^2\wh{p_n}
$$
into the asymptotics \er{asmnss} and \er{asmn}
we obtain \er{asmnss0} and \er{asmn0}.
\BBox

\section{Trace formula}
\setcounter{equation}{0}
\lb{Sect8}

\subsection{Asymptotics of the characteristic function}
Below we need the following result about the characteristic function.

\begin{lemma}
Let $p''',q''\in L^1(\T)$. Then
\[
\lb{assgmsh}
D(\l)={2e^{z({3\/2}-{p_0\/z^2})}\/3\sqrt3\l}
\Big(1+{V(0)\/\l}\Big)
\lt(\sin\Big({\sqrt3\/2}z+{p_0\/\sqrt3z}\Big)
+e^{{\sqrt3\/2}\Im z}O(z^{-4})\rt)
\]
as $|\l|\to\iy,\l\in\C_+,\Re\l\ge 0$, where $V=q-{p'\/3}$, see \er{deffh}.

\end{lemma}

\no{\bf Proof.} Let $p''',q''\in L^1(\T)$.
Let $|z|\to\iy,z\in \cZ_+$.
Integrating by parts in the asymptotics \er{asGljpr} we obtain
$$
\cB_{\ell j}(x,z)
=O(z^{-1}),
\qqq 1\le j,\ell=1,2,3,4,\qq j\ne \ell.
$$
Substituting these asymptotics into the definition \er{phi1jsc}
and using the definitions  \er{cK12}, \er{defW1ss} and  \er{defW2}
we obtain
$$
\begin{aligned}
\zeta_{1}(x,z)={\ol\vk p(x)\/3\sqrt3z^2}
+{V(x)\/3z^3}
+O(z^{-4}),
\\
\zeta_{2}(x,z)={\vk p(x)\/3\sqrt3z^2}
+{V(x)\/3z^3}+
O(z^{-4}),
\\
\zeta_{3}(x,z)={p(x)\/3z^2}
+{V(x)\/3z^3}+
O(z^{-4}),
\end{aligned}
$$
where $V(x)={1\/3}(h(x)+\ol h(x))=q(x)-{p'(x)\/3}$ and $h,\vk$
are given by \er{defh}.
Then the asymptotics \er{asphijepr} gives
$$
\phi_1(0,z)=\alpha(z)+O(z^{-4}),
\qq
\phi_1(1,z)=e^{z(\t-{2p_0\t^2\/3z^2})}
\big(\alpha(z)+O(z^{-4})\big),
$$
$$
\phi_2(0,z)=\beta(z)+O(z^{-4}),
\qq
\phi_2(1,z)=e^{z(\t^2-{2p_0\t\/3z^2})}
\big(\beta(z)+O(z^{-4})\big),
$$
\[
\lb{asphi3}
\phi_3(2,z)=e^{2z(1-{2p_0\/3z^2})}
\Big(1+{p(0)\/3z^2}+{V(0)\/3z^3}
+O(z^{-4})\Big),
\]
where
\[
\lb{defAzBa}
\alpha(z)=1+{\ol\vk p(0)\/3\sqrt3z^2}+{V(0)\/3z^3},\qqq
\beta(z)=1+{\vk p(0)\/3\sqrt3z^2}+{V(0)\/3z^3}.
\]

Let $|z|\to\iy,z\in \cZ_+^+$. Substituting these asymptotics into
the asymptotics \er{asdphiS} and using the definition \er{definA} we obtain
\[
\lb{asdetphipr}
\det\phi(z)
=\phi_3(2,z)e^{z(-{1\/2}+{p_0\/3z^2})}
\big(\x(z)+e^{{\sqrt3\/2}\Im z}O(z^{-4})\big).
\]
where
$$
\x(z)=
\det\ma
\alpha(z)&
\beta(z)\\
\alpha(z)e^{iz({\sqrt3\/2}+{p_0\/\sqrt3z^2})}
&\beta(z)e^{-iz({\sqrt3\/2}+{p_0\/\sqrt3z^2})}
\am
=-2i\alpha(z)\beta(z)\sin\Big({\sqrt3\/2}z+{p_0\/\sqrt3z}\Big).
$$
The definitions \er{defAzBa} imply
$$
\x(z)
=-2i\Big(1-{p(0)\/3z^2}
+{2V(0)\/3z^3}\Big)
\lt(\sin\Big({\sqrt3\/2}z+{p_0\/\sqrt3z}\Big)
+e^{{\sqrt3\/2}\Im z}O(z^{-4})\rt).
$$
Substituting this asymptotics and  \er{asphi3} into \er{asdetphipr} we obtain
\[
\lb{asdetphif}
\det\phi(z)
=-2ie^{z({3\/2}-{p_0\/z^2})}
\Big(1+{V(0)\/z^3}\Big)
\big(\sin\Big({\sqrt3\/2}z+{p_0\/\sqrt3z}\Big)+e^{{\sqrt3\/2}\Im z}O(z^{-4})\big).
\]

The asymptotics \er{4g.aswtGpr} and \er{asGjjpr} imply
$$
\det \cX(x,z)=1+\Tr \cB(x,z)+O(z^{-6})=1+O(z^{-4})\qq\forall\ \ x\in\R.
$$
Substituting these asymptotics and \er{detOm}
into \er{idPhi} we obtain
$$
\det A(0,z)=-i3\sqrt3z^3\big(1+O(z^{-4})\big).
$$
Substituting the last asymptotics and the asymptotics \er{asdetphif} into
the identity \er{idsigma} we obtain
the asymptotics \er{assgmsh}.
\BBox

\subsection{Trace formula}
We prove the trace formula for the operators $H_t,H$.

\medskip

\no {\bf Proof of Theorem~\ref{ThTrf}.}
Let  $p''',q''\in L^1(\T)$.
Introduce the function $F(t)=\sum_{n=-N}^N\m_n(t),t\in\T$,
where $N$ is given in Lemma \ref{CLev}~ii),
and the resolvents $R_t(\l)=(H_t-\l)^{-1}$.
Then we have
$$
\begin{aligned}
F(t_1)-F(t_2)=\fr{1}{2\pi i}\oint_{\G_N}\l
\Tr\big(R_{t_1}(\l)-R_{t_2}(\l)\big)d\l,
\\
\m_n(t_1)-\m_n(t_2)=\fr{1}{2\pi i}\oint_{\ell_n}\l
\Tr\big(R_{t_1}(\l)-R_{t_2}(\l)\big)d\l\qqq\forall\qq n>N,
\end{aligned}
$$
where the contours $\G_N$ and $\ell_n$ are given by
$$
\G_N=\Big\{\l\in\C:|\l|=\Big({2\pi (N+{1\/3})\/\sqrt3}\Big)^3\Big\},\qq
\ell_n=\Big\{\l\in\C:\Big|z-{2\pi n\/\sqrt3}\Big|=\fr{\pi}{4}\Big\}.
$$
Using the identities
$$
\begin{aligned}
R_{t_1}(\l)-R_{t_2}(\l)=
R_{t_1}(\l)\big(H_{t_2}-H_{t_1}\big)R_{t_2}(\l)
\\
=R_{t_1}(\l)\Big(\pa(p_{t_2}-p_{t_1})+(p_{t_2}-p_{t_1})\pa
+q_{t_2}-q_{t_1}\Big)R_{t_2}(\l),
\end{aligned}
$$
where $f_t=f(\cdot+t)$,
we obtain
$$
\big|F^{(k)}(t_1)-F^{(k)}(t_2)\big|\le
C\Big(\|p_{t_2}^{(k+1)}-p_{t_1}^{(k+1)}\|_\iy
+\|p_{t_2}^{(k)}-p_{t_1}^{(k)}\|_\iy
+\|q_{t_2}^{(k)}-q_{t_1}^{(k)}\|_\iy\Big)
$$
for some constant $C>0$ and $k=0,1$, where
$f^{(0)}=f,f^{(k)}=\fr{d^kf}{dt^k}$.
These estimates imply $F\in C^1(\T)$.
The similar arguments show that
$\m_n\in C^1(\T)$ for all $n>N$.

The asymptotics \er{asmn} shows that the series \er{trftr}
converges absolutely and uniformly in $t\in\T$.

Let $D(\l,t)=D(\l,p_t,q_t)$ and $D(\l)=D(\l,0)$.
The asymptotics \er{assig0} shows that $\log{D(\l,t)\/D(\l)}$
is well defined on the contours $\G_N$ for large $N\in\N$ by
the condition $\log 1=0$ and for $N\in\N$ large enough
we have
\[
\lb{contitr}
{1\/2\pi i}\oint_{\G_N}\l\ d\log{D(\l,t)\/D(\l)}=
{1\/2\pi i}\oint_{\G_N} \sum_{n\in\Z} \Big({\l\/\l-\m_n(t)}
-{\l\/\l-\m_n(0)}\Big)d\l
=\sum_{n=-N}^N\big(\m_n(t)-\m_n(0)\big).
\]
Let $\l\in \G_N,N\to\iy$.
The asymptotics \er{assgmsh} and the identity \er{ssymia} give
$$
{D(\l,t)\/D(\l)}
=1+{\sqrt3\/\l}\big(V(t)-V(0)\big)+O(z^{-4}),
$$
which yields
$$
\log{D(\l,t)\/D(\l)}={\sqrt3\/\l}\big(V(t)-V(0)\big)
+O(z^{-4}).
$$
Integrating by parts we obtain
$$
{1\/2\pi i}\oint_{\G_N}\l\ d\log{D(\l,t)\/D(\l)}=
-{1\/2\pi i}\oint_{\G_N}\log{D(\l,t)\/D(\l)}d\l.
$$
Using the estimate
$$
\lim_{N\to\iy}\oint_{\G_N}O(z^{-4})d\l=0
$$
we obtain
$$
\lim_{N\to\iy}{1\/2\pi i}\oint_{\G_N}\l\ d\log{D(\l,t)\/D(\l)}
=V(0)-V(t).
$$
The identity \er{contitr} gives the identity \er{trftr}.
\BBox

\medskip

\footnotesize

\no {\bf Acknowledgments.} \footnotesize A. Badanin was supported by
the RFBR grant  No 19-01-00094. E. Korotyaev was supported by the
RSF grant  No. 18-11-00032.

\end{document}